%% file: main.tex
\documentclass{llncs}

\usepackage[utf8]{inputenc}
\usepackage[T1]{fontenc}
\usepackage[english]{babel}

\usepackage{graphicx}
\usepackage{booktabs}
\usepackage{mathptmx}

\newcommand{\smrstovak}{\vspace{-1mm}}

\usepackage{amssymb,amsmath,amsfonts}
\usepackage{tikz,pgffor}

\usetikzlibrary{arrows}
\usetikzlibrary{shapes}
\usetikzlibrary{automata}
\usetikzlibrary{decorations.pathmorphing} 
\usetikzlibrary{decorations.pathreplacing} 
\usetikzlibrary{calc} 

\usepackage{txfonts}
\usepackage{color}
\usepackage{wrapfig}
\usepackage[linesnumbered,ruled]{algorithm2e}
\usepackage{float}
\usepackage{pgfplots}

\tikzstyle{max}=[thick,draw,minimum size=1.4em,inner sep=0em]
\tikzstyle{min}=[diamond,thick,draw,minimum size=1.4em,%
    inner sep=0em]
\tikzstyle{ran}=[circle,thick,draw,minimum size=1.4em,%
    inner sep=0em]
\tikzstyle{act}=[circle,thick,draw,fill,minimum size=.7em,%
    inner sep=0em]
\tikzstyle{mc}=[rounded corners,thick,draw,minimum size=1.4em,%
    inner sep=.5ex]
\tikzstyle{tran}=[thick,draw,->,>=stealth]
\tikzstyle{loop left}=[tran, to path={.. controls +(150:.5)
    and +(210:.5) .. (\tikztotarget) \tikztonodes}]
\tikzstyle{loop right}=[tran, to path={.. controls +(30:.5)
    and +(330:.5) .. (\tikztotarget) \tikztonodes}]
\tikzstyle{loop above}=[tran, to path={.. controls +(60:.5)
    and +(120:.5) .. (\tikztotarget) \tikztonodes}]
\tikzstyle{loop below}=[tran, to path={.. controls +(240:.5)
    and +(300:.5) .. (\tikztotarget) \tikztonodes}]

\newcommand{\theoremlike}[2]{\par\medskip\penalty-250\refstepcounter{theorem}{{\bfseries\noindent#2
\ref{#1}.}}}

\newcommand{\thmhelperpre}[2]{\theoremlike{#1}{#2}}
\newcommand{\thmhelperpost}{\par\medskip}

\newcommand{\Nset}{\mathbb{N}}

\newcommand{\Qset}{\mathbb{Q}}
\newcommand{\Rset}{\mathbb{R}}

\newcommand{\Rsetpo}{\mathbb{R}_{\ge 0}}

\newcommand{\val}{\mathnormal{remain}}

\newcommand{\de}[1]{\mathit{d#1}}  

\newcommand{\ctmc}{\mathcal{C}}
\newcommand{\dCTMC}{d-CTMC}
\newcommand{\delay}{\mathnormal{delay}}
\newcommand{\algone}{\texttt{IPH-shift} }
\newcommand{\algtwo}{\texttt{IPH-slice} }
\newcommand{\algonep}[1]{\texttt{IPH-shift[#1]}}
\newcommand{\algtwop}[1]{\texttt{IPH-slice[#1]}}
\newcommand{\dctmc}{\mathcal{D}}
\newcommand{\events}{\mathcal{E}}
\newcommand{\states}{S}
\newcommand{\sched}{\mathbf{E}}
\newcommand{\occur}{\mathrm{Succ}}

\newcommand{\dist}{\mathcal{D}}
\newcommand{\distribution}{\alpha}

\newcommand{\figpath}{figures}
\newcommand{\init}{\textsf{init}}
\newcommand{\sent}{\textsf{sent}}
\newcommand{\lost}{\textsf{lost}}
\newcommand{\ok}{\textsf{ok}}
\newcommand{\start}{\textsf{s}}
\newcommand{\aone}{$\mathsf{a}_1$}
\newcommand{\atwo}{$\mathsf{a}_2$}
\newcommand{\tone}{$\mathsf{t}_1$}
\newcommand{\ttwo}{$\mathsf{t}_2$}
\newcommand{\uone}{\mathsf{u}_1}
\newcommand{\utwo}{\mathsf{u}_2}
\newcommand{\uthree}{\mathsf{u}_3}
\newcommand{\ufour}{\mathsf{u}_4}

\newcommand{\send}{\textsf{send}}
\newcommand{\ack}{\textsf{ack}}
\newcommand{\err}{\textsf{err}}
\newcommand{\timeout}{\textsf{timeout}}
\newcommand{\discrete}{\mathsf{d}}

\newcommand{\add}{Err}

\newif\iffig
\figtrue

\begin{document}

\title{Dealing with Zero Density Using \\ Piecewise Phase-type Approximation
\thanks{
This work is supported by the EU 7th Framework Programme under grant agreements 295261 (MEALS) and 318490 (SENSATION), Czech Science
Foundation grant No.~P202/12/G612 [aktualizovat], the DFG Transregional
Collaborative Research Centre SFB/TR 14 AVACS, and by the CAS/SAFEA
International Partnership Program for Creative Research Teams. 
}
}

\author{ \v{L}ubo\v{s} Koren\v{c}iak\inst{1} \and Jan Kr\v{c}\'al\inst{2} \and
  Vojt\v{e}ch~{\v{R}}eh\'ak\inst{1}}

\institute{
  Faculty of Informatics, Masaryk University,
  Brno, Czech Republic\\
  {\{korenciak,\,rehak\}\!@fi.muni.cz}
  \and 
  Saarland University -- Computer Science, Saarbr\"ucken, Germany\\
  {krcal\!@cs.uni-saarland.de}
  }

\maketitle

\begin{abstract}

Every probability distribution can be approximated
up to a given precision by a phase-type distribution, i.e.\ a distribution encoded by a continuous time Markov chain (CTMC). 
However, an excessive number of states in the corresponding CTMC is needed for some standard distributions, in particular most distributions with regions of zero density such as uniform or shifted distributions.
Addressing this class of distributions, we suggest an alternative representation by CTMC extended with \emph{discrete-time} transitions. 
Using discrete-time transitions we split the density function into multiple intervals. Within each interval, we then approximate the density with standard phase-type fitting. 
We provide an experimental evidence that our method requires only a moderate number of states to approximate such distributions with regions of zero density.
Furthermore, the usage of CTMC with discrete-time transitions is supported by 
a number of techniques for their analysis.
%
%
Thus, our results promise an efficient approach to the transient analysis of a class of non-Markovian models.

\end{abstract}

\input{intro}
\input{prelim}

\input{iph}

\input{methods}

\input{results}
\section{Conclusions and Future Work}

In this paper we introduced an alternative approach to phase-type approximation of non-Markovian models, called Interval phase-type approximation. Instead of producing a CTMC, our method approximates the original model using a d-CTMC. The method provides substantial reduction of the state space which may lead to a lower analysis time as indicated by our experiments.

There are several directions for future work. First, a comparison
of the existing algorithms~\cite{guet2012delayed,lindemann1999transient,horvath2012transient} 
  that can be applied to the transient analysis of d-CTMC would be highly welcome. 
Furthermore, for the best algorithm for \dCTMC, one can perform a more detailed comparison of its running times on the \dCTMC{} obtained by the IPH fitting with the analysis times of other available methods (such as the standard PH fitting). 
%
Second, we believe that further heuristics can increase the efficiency of IPH or its applicability to a wider class of distributions. 
Finally, our method justifies the importance of research on 
%
further analysis algorithms for \dCTMC.
%

\paragraph{Acknowledgement} 
We would like to thank Vojt\v{e}ch Forejt, Andr\'as Horv\'ath, David Parker, and Enrico Vicario for inspiring discussions.


\bibliographystyle{plain}
\bibliography{str-short,concur,nase}

\newpage
\appendix
\input{app}

\end{document}

%% file: intro.tex

\section{Introduction}
\label{sec-intro}

In the area of performance evaluation and probabilistic verification, discrete-event systems (DES) are a prominent modelling formalism. It includes models such as continuous-time Markov chains, stochastic Petri nets, or generalized semi-Markov processes. A DES is a random process that is initialized in some state 
and then moves from state to state in continuous-time whenever an event occurs. 
Every time a state is entered, some of the events get \emph{initiated}. An initiated event then occurs after a delay chosen randomly according to its distribution function.
%
When no restrictions on the distribution functions are imposed, analysis of these models is complicated~\cite{brazdil2011fixed,horvath2012transient}, one often resorts to simulation~\cite{haas2002stochastic}. When all the distributions $F_e$ are exponential, the DES is then called a continuous-time Markov chain (CTMC) for which many efficient analysis methods exist~\cite{jensen1953markoff,baier2003model} thanks to the memoryless property of the exponential distribution.
%
%
%
Hence, an important method for analysing DES is to \emph{approximate} it by a CTMC using \emph{phase-type} (PH) approximation. Roughly speaking, each event $e$ such that its distribution function is not exponential is replaced by a small CTMC $\ctmc_e$. This CTMC has 
%
a designated absorbing state such that the time it takes to reach the absorbing state is distributed as closely as possible to the given distribution function. 
A well known result~\cite{Neuts81} states that any
continuous probability distribution can be fitted up to a given precision by the PH approximation. 
Nevertheless, the closer the approximation, the more states it requires in the CTMC. For some lower bounds on the number of
required states see, e.g.,~\cite{AS87Erlang,cinneide1999open_problems,Faddy98InferringCox,fackrell2005fitting}
.

\begin{figure}[t]
  \centering
\iffig 
\input{figures/figure1.pgf}
\fi
\caption{Three usages of the discrete-time events for PH approximation. In the figures there are the densities (with thick 
grey lines), their standard PH approximations with $4$ and $40$ phases, and their IPH approximation with $30$ phases.
On the left, the discrete-time event $d$ postpones the start of the CTMC $\ctmc$ fitted to the area of positive density. 
On the right, the discrete-time event can be used directly, instead of its continuous approximation.
In the middle, $3$ discrete-time events split the support into $4$ intervals with different approximations $\ctmc_1$, $\ctmc_2$, $\ctmc_3$, and $\ctmc_4$. 
Note that for a distribution with a steep change in density at its upper bound (such as the uniform distribution), PH fitting performs well on the first \emph{half} of the support; logarithmic partitioning into intervals works better than equidistant.
}
\label{fig:method-illustration}
\end{figure}
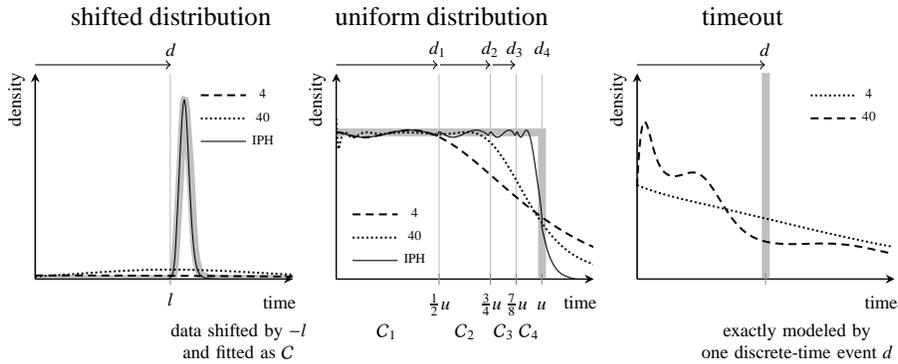

In this paper we propose another approach for approximating probability distributions where phase-type requires extreme amount of states to be fitted precisely~\cite{cinneide1999open_problems,fackrell2005fitting}. 
%
%
In particular, we deal with distributions often encountered in practice that we call \emph{interval distributions} and that are supported on a proper subinterval of $[0,\infty)$. For example distributions of events that cannot occur before time $l>0$ such as due to physical limits when \emph{sending a packet}; or that 
cannot occur after time $u < \infty$ such as \emph{waiting for a random amount of time} in a collision avoidance protocol; or that occur exactly after time $l=u$ such as \emph{timeouts}.
%
%
%
We address these interval distributions by an approach that we call 
\emph{Interval phase-type (IPH)} approximation.
The crucial point is that it allows to separate the discrete and the continuous nature of these distributions by enriching the output formalism. 
Along with the exponential distribution of the CTMC we allow discrete-time events (also called fixed-delay, deterministic, or timeout events) and denote it as \dCTMC{}.\footnote{Note that the formalism of \dCTMC{} is inspired by the previously studied similar formalisms of \emph{deterministic and stochastic Petri nets}~\cite{marsan1987petri} and \emph{delayed CTMC} \cite{guet2012delayed}.} 
As illustrated in Figure~\ref{fig:method-illustration}, the usage of discrete-time events for approximating a non-exponential distribution is threefold:
\begin{enumerate}
 \item For an event $e$ with occurrence time bounded from below by $l > 0$, an occurrence of a discrete-time event $d$ splits the waiting into two parts -- an initial part of length $l$ where the event $e$ cannot occur and the rest that can be more efficiently approximated by a CTMC $\ctmc$ using standard PH methods. 
 \item For an event $e$ with occurrence time bounded from above by $u < \infty$, a series of discrete-time events partition the support of its distribution into $n$ subintervals. 
The system starts in the chain $\ctmc_1$ which is the standard PH approximation of the whole density. 
In parallel to movement in $\ctmc_1$ a discrete-time event $d_1$ is awaited with its occurrence set to the beginning of the second interval. If the absorbing state in $\ctmc_1$ is not reached before $d_1$ occurs, the system moves to $\ctmc_2$. The chain $\ctmc_2$ is fitted to the whole remaining density conditioned by the fact that the event does not occur before the beginning of the second interval. Similarly, another discrete-time event $d_2$ is awaited in $\ctmc_2$ with its occurrence set to the beginning of the third interval, etc. 
The last interval is not ended by any discrete-time event; occurrence of the event $e$ thus corresponds to reaching any absorbing state in any of $\ctmc_1, \ldots, \ctmc_n$.
\item An event with constant occurrence time ($\ell =u$) is directly a discrete-time event.
\end{enumerate}
%

\paragraph{Example} As our running example, we consider the Alternating bit protocol. 
Via a lossy FIFO channel, a transmitter attempts to send a sequence of messages, each endowed with a one-bit sequence number -- alternating between $0$ and $1$. 
The transmitter keeps resending each message 
until it is acknowledged by its sequence number (the receiver sends back the sequence number of each incoming message).
%
%
%
As resending of messages is triggered by a timeout, setting an appropriate value for the timeout is essential in balancing the performance of the protocol and the network congestion.
For a given timeout, one may ask, e.g., 
\emph{what is the probability that $10$ messages will be successfully sent in $100 ms$?} In the next section we show a simple DES model of this protocol. Subsequently, we show the CTMC model yielded by a PH approximation of individual events, and the {\dCTMC} model obtained by our proposed IPH approximation.


\paragraph{Our contribution}

We propose an alternative approach to PH approximation, resulting in a CTMC enriched with fixed-delay events. Our approach is tailored to interval probability distributions that are often found in reality and for which the standard continuous PH approximation requires a substantial amount of states.
We performed an experimental evaluation of our approach.
In the evaluation, we represent (1) the lower-bounded distributions by the distribution of the \emph{transport time in network communication} and (2) the upper-bounded distributions by the \emph{uniform} distribution.
%
For both cases, we show that our approach requires only a moderate number of states to approximate these distributions up to a given error. Thus, for DES models with interval distributions our approach promises a viable method for transient analysis as also indicated by our experiments.

\paragraph{Related work}
Already in the original paper of Neuts~\cite{Neuts81}, the fixed-delay and
shifted exponential distributions have been found difficult to
fit with a phase-type approximation. This fact was explicitly
quantified by Aldous and Shepp \cite{AS87Erlang} showing that the Erlang
distribution is the best PH fitting for the fixed-delay distributions.
A notoriously difficult example of a shifted distribution is the data set measuring the length of
eruptions of a geyser in the Yellowstone National Park \cite{silverman1986density} whose PH approximation
has been discussed in, e.g., \cite{asmussen1996fitting,Faddy98InferringCox}.
Also heavy tailed distributions often found in telecommunication systems are hard to fit; similarly to our method, separate fitting of the body and the tail of such distributions is used~\cite{FW98Heavytail,HT-Performance-02}.
%
%

Apart from continuous PH fitting, there are several other methods applicable to analysis of DES with interval distributions. 
First, there are several symbolical solution methods ~\cite{alur2006bounded,bernadsky2007symbolic,horvath2011probabilistic,horvath2012transient} for direct analysis of DES with non-exponential events.
Usually, \emph{expolynomial} distributions are allowed; non-expolynomial distributions need to be fitted by expolynomials -- a problem far less studied than standard PH fitting. Our approach can be understood as a specific fitting technique that uses a limited subclass of expolynomial distributions (resulting in models with a wider range of analysis techniques).
%
%
%
Second, interval distributions can be efficiently fitted by discrete phase-type approximation~\cite{bobbio2003acyclic}. Instead of a CTMC, this method yields a \emph{discrete-time} Markov chain
(DTMC) where each discrete step corresponds to elapsing some fixed $\delta$ time units. 
Note however that this method usually requires to discretize \emph{all the events} of a DES into a DTMC. 
To analyse faithfully a DES with many parallel events one either needs to use a very small $\delta$~\cite{zhang2010model} or to allow occurrence of multiple events within each $\delta$-time step~\cite{molloy1985discrete,hatefi2013improving}, exponentially increasing the amount of states or transitions in the DTMC, respectively.
Third, similarly to our approach, ideas for combining discrete PH approximation with continuous PH approximation have already appeared~\cite{jones2001phased,haddad2005performance}. To the best of our knowledge, no previous work considers combining these two approaches on \emph{one} distribution having both discrete and continuous ``nature''. Expressing the continuous part of such a distribution using continuous PH again decreases the coincidence of parallel discrete events discussed above.
Note that with {\dCTMC}, one can freely combine continuous PH, discrete PH, and interval PH for approximation of different events of a DES.
%

\smrstovak
\paragraph{Organization of the paper}

In Section~\ref{sec-prelim}, we define the necessary preliminaries. In Section~\ref{sec-IPH}, we describe the IPH approximation method and briefly review the analysis techniques for d-CTMC. The paper is concluded by an experimental evaluation in Section~\ref{sec-results}.

%% file: figures/figure1.pgf
\begin{tikzpicture}[font=\scriptsize]

\begin{scope}[shift={(-4,0)}] 

\draw[->,] (0,2.85) -- node[above=-0.1,pos=1] {$d$}  (1.8,2.85);
\draw[->,draw=none] (1.8,2.85) --  node[below=95,pos=0.5,text width=2cm,text centered] {data shifted by $-l$ \\ and fitted as $\mathcal{C}$} (3.6,2.85);
\begin{axis}[
	ymin=0,ymax=12,xmin=3, xmax=5,
	no markers,
	axis y line=left,axis x line=bottom,
 	xtick=\empty, ytick=\empty,
	width=5cm
]

\addplot[line width=3pt, lightgray,smooth] table[x=x,y=ping] {\figpath/figure1.data};

\end{axis}
\begin{axis}[
	restrict x to domain=3:5,
	ymin=0,ymax=12,
	no markers,
	axis y line=left,axis x line=bottom,
	extra x ticks={4.05},
	extra x tick style={grid=major},
	extra x tick labels={$l$},
 	xtick=\empty, ytick=\empty,
	width=5cm,
 	xlabel=time, ylabel=density,
 	xlabel style={shift={(0.45,0.15)}},ylabel style={shift={(0.3,-0.36)}},
	legend style={font=\tiny,draw=none,fill=none}
]

\addplot[line width=0.75pt,black,smooth, densely dashed] table[x=x,y=4] {\figpath/figure1.data};
\addplot[line width=0.75pt,black,smooth,densely dotted] table[x=x,y=40] {\figpath/figure1.data};
\addplot[black,smooth] file {\figpath/tikz_iph.pdf.txt};

\addlegendentry{4}
\addlegendentry{40}
\addlegendentry{IPH}

\end{axis}
\end{scope}

\begin{scope}[shift={(0,0)}] 

\node  at (-2.2,3.5) {\normalsize shifted distribution};
\node  at (1.4,3.5) {\normalsize uniform distribution};
\node  at (5.4,3.5) {\normalsize timeout};

\draw[->,] (0,2.85) -- node[above=-0.1,pos=1] {$d_1$} node[below=95,pos=0.5] {$\mathcal{C}_1$} (1.35,2.85);
\draw[->,] (1.38,2.85) -- node[above=-0.1,pos=1] {$d_2$} node[below=95,pos=0.5] {$\mathcal{C}_2$} (2.05,2.85);
\draw[->,] (2.08,2.85) -- node[above=-0.1,pos=1] {$d_3$} node[below=95,pos=0.5] {$\mathcal{C}_3$} (2.38,2.85);
\draw[draw=none] (2.38,2.85) -- node[above=-0.1,pos=1] {$d_4$} node[below=95,pos=0.5] {$\mathcal{C}_4$} (2.75,2.85);

\begin{axis}[
	ymin=0,ymax=0.7,xmin=0, xmax=2.5,
	no markers,
	axis y line=left,axis x line=bottom,
 	xtick=\empty, ytick=\empty,
	width=5cm,
]

\addplot[line width=3pt, lightgray] plot coordinates {(0,0.5) (2,0.5) (2,0)};

\end{axis}
\begin{axis}[
	ymin=0,ymax=0.7,xmin=0, xmax=2.5,
	no markers,
	axis y line=left,axis x line=bottom,
	extra x ticks={1,1.5,1.75,2},
	extra x tick style={grid=major},
	extra x tick labels={$\frac{1}{2}u$,$\frac{3}{4}u$,$\frac{7}{8}u$,$\vphantom{\frac{1}{2}}u$},
 	xtick=\empty, ytick=\empty,
	width=5cm,
 	xlabel=time, ylabel=density,
 	xlabel style={shift={(0.45,0.15)}},ylabel style={shift={(0.3,-0.36)}},
	legend style={font=\tiny,draw=none,fill=none,at={(0.4,0.4)}}
]

\addplot[line width=0.75pt,black,densely dashed] table[x=x,y=four] {\figpath/figure6.data};
\addplot[line width=0.75pt,black,smooth,densely dotted] table[x=x,y=fourty] {\figpath/figure6.data};
\addplot[black,smooth] table[x=x,y=rescaled] {\figpath/figure6.data};

\addlegendentry{4}
\addlegendentry{40}
\addlegendentry{IPH}

\end{axis}
\end{scope}

\begin{scope}[shift={(4,0)}] 

\draw[->,] (0,2.85) -- node[above=-0.1,pos=1] {$d$} (1.7,2.85);
\draw[->,draw=none] (0,2.85) --  node[below=95,pos=0.5,text width=4cm,text centered] {exactly modeled by \\ one discrete-time event $d$} (4.25,2.85);
\begin{axis}[
	ymin=0,ymax=0.13,xmin=0, xmax=20,
	no markers,
	axis y line=left,axis x line=bottom,
 	xtick=\empty, ytick=\empty,
	width=5cm
]

\addplot[line width=3pt, lightgray] plot coordinates {(10,0) (10,1)};
;
\end{axis}

\begin{axis}[
	ymin=0,ymax=0.13,xmin=0, xmax=20,
	no markers,
	axis y line=left,axis x line=bottom,
	extra x ticks={10},
	extra x tick style={grid=major},
	extra x tick labels={},
 	xtick=\empty, ytick=\empty,
	width=5cm,
 	xlabel=time, ylabel=density,
 	xlabel style={shift={(0.45,0.15)}},ylabel style={shift={(0.3,-0.36)}},
	legend style={font=\tiny,draw=none,fill=none}
]

%
%
%
%

  \addplot[line width=0.75pt,black,densely dotted] table[x=x,y=40] {\figpath/figure1_timeout.data};
  \addplot[line width=0.75pt,black, densely dashed] table[x=x,y=4] {\figpath/figure1_timeout.data};

%
%
%
%
 \addlegendentry{4}
 \addlegendentry{40}

%
%
%
%
\end{axis}
\end{scope}

\end{tikzpicture}

%% file: prelim.tex
\section{Preliminaries}
\label{sec-prelim}

\smrstovak
We denote by $\Nset$, $\Qset$, and $\Rset$ the sets of
natural, rational, and real
numbers, respectively.
For a finite set $X$, $\dist(X)$ denotes the set of all discrete probability distributions over $X$. 

\smrstovak\smrstovak
\subsubsection{Modelling formalisms}
There are several equivalent formalisations of DES. Here we define \emph{generalized-semi Markov processes} that contain both CTMC and \dCTMC{} as subclasses. 
%
Let $\events$ be a finite set of \emph{events} where each event is either a \emph{discrete-time} event or a \emph{continuous-time event}. To each discrete-time event $e$ we assign its delay $\delay(e) \in \Qset$.
To each continuous-time event $e$ we assign a \emph{probability density
  function} $f_e : \Rset \rightarrow \Rset$ such that $\int_{0}^{\infty}
f_e(x)\, \de{x} = 1$. An event is called \emph{exponential} if it is a
continuous-time event with density function $f(x) = \lambda \cdot
e^{-x\lambda}$ where $\lambda > 0$ is its rate.

\begin{definition}
  A \emph{generalized semi-Markov process} (GSMP) is a tuple 
  $(\states,\events,\sched,\occur,\distribution_0)$ where 
\begin{itemize}
 \item $\states$ is a finite set of \emph{states}, 
 \item $\events$ is a finite set of \emph{events},
 \item $\sched : \states \rightarrow 2^{\events}$ assigns to each state $s$ a set of events 
  \emph{active} in $s$,
 \item $\occur : \states \times \events \rightarrow \dist(\states)$ is the \emph{successor} function, i.e. 
  it assigns a probability distribution specifying the successor state
  to each state and event that occurs there,
 \item $\distribution_0 \in \dist(\states)$ is the \emph{initial distribution}.
\end{itemize}
We say that a GSMP is a \emph{continuous-time Markov chain} (CTMC) if every events of $\events$ is exponential. We say that a GSMP is a \emph{continuous-time Markov chain with discrete-time events} (\dCTMC{}) if every event of $\events$ is either exponential or discrete-time.
\end{definition}

%
The run of a GSMP starts in a state $s$ chosen randomly according to
$\distribution_0$. At start, each event $e \in \sched(s)$ is
\emph{initialized}, i.e. the amount of time $\val(e)$ remaining until it
occurs is (1) set to $\delay(e)$ if $e$ is a discrete-time event, or (2) chosen randomly according to the density function $f_e$ if $e$ is a continuous-time event. 
Let the process be in a state $s$ and let the event $e$ have the minimal remaining time $t = \val(e)$ among all events active in $s$. 
The process waits in $s$ for time $t$ until the event $e$ occurs, then the next state $s'$ is chosen according to the distribution $\occur(s,e)$%
\footnote{For the sake of simplicity, when multiple events $X = \{e_1,\ldots,e_n\}$ occur simultaneously, the successor is determined by the minimal element of $X$ according to some fixed total order on $\events$. A more general definition~\cite{brazdil2011fixed} allows to specify different behaviour for simultaneous occurrence of any subset of events.}.
Upon this transition, the remaining time of each event of $\sched(s)\smallsetminus \sched(s')$ which is not active any more is discarded, and each event of $\sched(s')\smallsetminus \sched(s)$ is initialized as explained above. Furthermore if the just occurred event $e$ belongs to $\sched(s')$, it is also initialized. For a formal definition we refer to~\cite{brazdil2011fixed}.
%

%

\begin{figure}[t]
  \centering
\iffig \input{figures/figure2.pgf} \fi
\caption{On the left, there is a GSMP model of sending \emph{a single} message using the the Alternating bit protocol. The set of events active in a state corresponds to the edges outgoing from that state. The event \textsf{timeout} is discrete-time with delay $10$~ms, \textsf{send} is exponential with rate $2$ meaning that it takes $0.5$~ms on average to send a message, \textsf{err} is exponential with rate $0.01$ corresponding to a packet being lost each $100$~ms of network traffic on average, and \textsf{ack} is continuous-time with density displayed in Figure~\ref{fig:method-shifted} on the left. In the middle, there are 2-phase PH approximations of events \textsf{ack} and \textsf{timeout}. On the right, there is a PH approximation of the GSMP model obtained roughly speaking as a product of the GSMP and the two PH components.}
\label{fig:example-gsmp}
\end{figure}
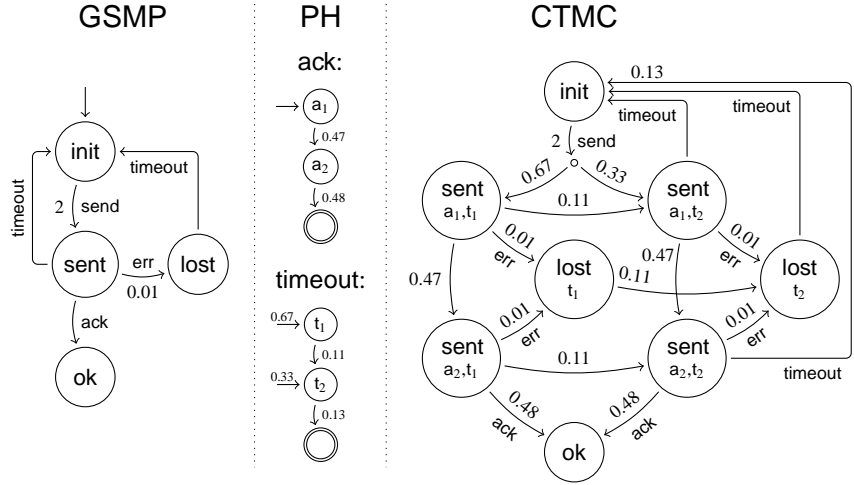

\paragraph{Example (continued)} 
To illustrate the definition, Figure~\ref{fig:example-gsmp} shows on the left a simplified GSMP model of the Alternating bit protocol. 
The transmitter sending a message corresponds to the exponential event
\textsf{send}. The whole remaining process of the message being transported
to the receiver, the receiver sending an acknowledgement message and the
acknowledgement message being transported back to the transmitter is
modelled using one continuous-time event \textsf{ack}. In parallel with the
event \textsf{ack}, there is a discrete-time event \textsf{timeout} and an
exponential event \textsf{err} 
representing a packet loss.
%

To exemplify the semantics, assume the process is in the state \textsf{sent} with $\val(\mathsf{timeout}) = 10$, $\val(\mathsf{ack})$ is chosen randomly to $12.6$ and $\val(\mathsf{err})$ is chosen randomly to $7.2$. Hence, after $7.2$ time units the event \textsf{err} occurs and the process moves to the state \textsf{lost} with $\val(\mathsf{timeout}) = 2.8$. After further $2.8$ time units, the timeout elapses and the process moves to the state \textsf{init} where $\val(\textsf{send})$ is chosen randomly to $0.8$. After this time, the process moves to \textsf{send} where $\val(\mathsf{timeout})$ is again set to $10$ and $\val(\mathsf{ack})$ and $\val(\mathsf{err})$ are again sampled 
according to their densities and so on. 
In the next section, we show the PH approximation of this model.


\subsubsection{Continuous PH approximation} 
Continuous PH can be viewed as a class of algorithms
\begin{itemize}
 \item which take as input the number of phases $n \in \Nset$ and a probability density function $f$ of a positive random variable%
, and
 \item output a CTMC $\ctmc$ with states $\{0,1,\ldots,n\}$ where $0$ is an absorbing\footnote{We say that a state $s$ is \emph{absorbing} if there are no outgoing transitions, i.e. $\sched(s) = \emptyset$.} state.
\end{itemize}

\noindent
Any such CTMC $\ctmc$ defines a positive random variable $X$ expressing the time when the absorbing state $0$ is reached in $\ctmc$. 
Let $\hat{f}$ denote the probability density function of 
$X$.
%
A possible goal of a PH algorithm is to minimize the \emph{absolute density
  difference} \cite{bobbio1994benchmark}\footnote{
Note that there are PH methods that do not allow specifying the number of phases. For further metrics for evaluating quality of PH approximation, see, e.g.,~\cite{bobbio1994benchmark}.
  }
\begin{align*}
 \int_0^\infty \lvert f(x) - \hat{f}(x) \rvert \; \de{x}. \tag{\add}
\end{align*}
%
%

\paragraph{Example (continued)}
When building a CTMC model of the Alternating bit protocol from the GSMP model, we need to approximate the non-exponential events \textsf{ack} and \textsf{timeout}. Their simple approximation and the whole CTMC model of the system is depicted in Figure~\ref{fig:example-gsmp} on the right. Observe that each state of the whole model needs to be enriched with the phase-number of every non-exponential event scheduled in this state. The events are then defined in a natural way on this product state space.

\paragraph{}
In the next section we describe our extension of PH fitting with discrete-time events.

%
%
%
%
%
%


%% file: figures/figure2.pgf
\begin{tikzpicture}[font=\scriptsize]

\tikzstyle{state} = [draw, circle, minimum size=2.5em, outer sep=0.2em,font=\footnotesize]
\tikzstyle{smallstate} = [state, minimum size=1.3em,font=\scriptsize, inner sep=0.2em]
\tikzstyle{trans} = [->]
\tikzstyle{small} = [font=\tiny]
\tikzstyle{multi} = [text width=15,text centered, font=\scriptsize]

\begin{scope}[shift={(1.5,0)}] 
	\node[font=\large] at (-0.5,-0.5) {\textsf{GSMP}};
	
	\begin{scope}[shift={(0,-2.3)}] 
		\node[state] (init) at (-1,-0) {\init};
		\node[state] (sent) at (-1,-1.5) {\sent};
		\node[state] (lost) at (0.5,-1.5) {\lost};
		\node[state] (ok) at (-1,-3) {\ok};
		
		\path  
			(init) edge[trans,bend right=15] node[auto] {\send}node[auto,swap] {2} (sent)
			(sent) edge[trans,bend right=15] node[auto] {\ack} (ok)
			(sent) edge[trans,bend right=15] node[auto] {\err} node[auto,swap] {0.01} (lost);
		\draw [rounded corners=2,trans] (lost) |- node[auto,below,pos=0.75] {\timeout}  (init);
		\draw [rounded corners=2,trans] (sent.west) -- ($(sent.west)+(-0.2,0)$) |- node[auto,above,sloped,pos=0.25] {\timeout}  (init.west);
		\draw [trans] ($(init.north)+(0,0.4)$) -- (init);
	\end{scope}
\end{scope}

\begin{scope}[shift={(2.75,0)}] 
	\draw [dotted] (0,0) -- (0,-6.5);
\end{scope}

\begin{scope}[shift={(3.63,0)}] 
	\node[font=\large] at (0,-0.5) {\textsf{PH}};

	\begin{scope}[shift={(0,-1.1)}] 
		\node[font=\normalsize] at (0,0) {\ack:};
		\node[smallstate] (aone) at (0,-0.6) {\aone};
		\node[smallstate] (atwo) at (0,-1.4) {\atwo};
		\node[smallstate,accepting,outer sep=0.2em] (athree) at (0,-2.2) {};

		\path  
			(aone) edge[trans,bend right=15] node[auto,small] {$0.47$} (atwo)
			(atwo) edge[trans,bend right=15] node[auto,small] {$0.48$} (athree)
		;
		\draw [trans] ($(aone.west)+(-0.3,0)$) -- (aone);
	\end{scope}

	\begin{scope}[shift={(0,-4)}] 
		\node[font=\normalsize] at (0,0) {\timeout:};
		\node[smallstate] (tone) at (0,-0.6) {\tone};
		\node[smallstate] (ttwo) at (0,-1.4) {\ttwo};
		\node[smallstate,accepting,outer sep=0.2em] (tthree) at (0,-2.2) {};

		\path  
			(tone) edge[trans,bend right=15] node[auto,small] {$0.11$} (ttwo)
			(ttwo) edge[trans,bend right=15] node[auto,small] {$0.13$} (tthree)
		;
		\draw [trans] ($(tone.west)+(-0.3,0)$) node[above=-2, small,xshift=2] {$0.67$}  -- (tone);
		\draw [trans] ($(ttwo.west)+(-0.3,0)$) node[above=-2, small,xshift=2] {$0.33$}  -- (ttwo);
	\end{scope}
\end{scope}

\begin{scope}[shift={(4.5,0)}] 
	\draw [dotted] (0,0) -- (0,-6.5);
\end{scope}

\begin{scope}[shift={(7,0)}] 
	\node[font=\large] at (0,-0.5) {\textsf{CTMC}};
	
	\begin{scope}[shift={(0,-1.7)}] 
		\node[state] (cinit) at (0,0.2) {\init};
		\node[state, multi] (csent11) at (-1.5,-1.25) {{\footnotesize \sent} \\ \aone,\tone};
		\node[state, multi] (csent12) at (1.5,-1.25) {{\footnotesize \sent} \\ \aone,\ttwo};
		\node[state, multi] (csent21) at (-1.5,-3.35) {{\footnotesize \sent} \\ \atwo,\tone};
		\node[state, multi] (csent22) at (1.5,-3.35) {{\footnotesize \sent} \\ \atwo,\ttwo};
		\node[state, multi] (clost1) at (0,-2.3) {{\footnotesize \lost} \\ \tone};
		\node[state, multi] (clost2) at (3,-2.3) {{\footnotesize \lost} \\ \ttwo};
		\node[state] (cok) at (0,-4.6) {\ok};
		\node[draw,circle, inner sep=0.1em,outer sep=0.2em] (prob) at (0,-0.75) {};
		
		\path  
			(cinit.-95) edge[trans,bend right=15] node[auto] {\send} node[auto,swap] {2} (prob)
			(prob) edge[trans,bend left=15] node[auto,sloped,above,pos=0.4] {$0.67$} (csent11)
			(prob) edge[trans,bend right=15] node[auto,sloped,above,pos=0.4] {$0.33$} (csent12)
			(csent11) edge[trans, bend right=10] node[above,sloped] {$0.11$} (csent12)			
			(csent21) edge[trans, bend right=10] node[above,sloped] {$0.11$} (csent22)			
			(csent11) edge[trans, bend right=10] node[auto,swap] {$0.47$} (csent21)			
			(csent12) edge[trans, bend right=10] node[auto,swap,pos=0.15,left=-2] {$0.47$} (csent22)
			(csent11) edge[trans, bend right=10] node[above,sloped] {$0.01$} node[below,sloped] {\err} (clost1)
			(csent21) edge[trans, bend right=10] node[above,sloped] {$0.01$} node[below,sloped] {\err} (clost1)
			(csent12) edge[trans, bend right=10] node[above,sloped] {$0.01$} node[below,sloped] {\err} (clost2)
			(csent22) edge[trans, bend right=10] node[above,sloped] {$0.01$} node[below,sloped] {\err} (clost2)
			(clost1) edge[trans, bend right=10] node[pos=0.1,sloped,above] {$0.11$} (clost2)			
			(csent21) edge[trans, bend right=10] node[above,sloped] {$0.48$} node[below,sloped] {\ack} (cok)
			(csent22) edge[trans, bend left=10] node[above,sloped] {$0.48$} node[below,sloped] {\ack} (cok)
		;
		\draw [rounded corners=2,trans] (csent12) |- node[auto,below,sloped,pos=0.75] {\timeout} node[auto,above=5,sloped,pos=0.75] {0.13}   (cinit.-15);
		\draw [rounded corners=2,trans] (clost2) |- node[auto,below,sloped,pos=0.6] {\timeout}  (cinit);
		\draw [rounded corners=2,trans] (csent22.east) -- node[auto,below,sloped,pos=0.68] {\timeout}  ($(csent22.east)+(+1.6,0)$) |- (cinit.15);
	\end{scope}
\end{scope}

\end{tikzpicture}

%% file: iph.tex
\section{Interval Phase-type Approximation}
\label{sec-IPH}

The Interval phase-type (IPH) approximation addresses the interval probability distributions which are supported on a proper subinterval of $[0,\infty)$. Similarly as above,
\begin{itemize}
 \item it takes as input the number of phases $n \in \Nset$ and a probability density function $f$ of a positive random variable, and
 \item outputs a \dCTMC{} $\dctmc$ with states $\{0,1,\ldots,n\}$  where $0$ is absorbing.
\end{itemize}
The goal is again to minimize 
(\add) for $\hat{f}$ being the probability density function\footnote{For the error metrics (\add) we assume that the algorithm outputs a \dCTMC{} such that $X$ has a density (which holds for our algorithms presented later).} of the random variable expressing the time when the absorbing state $0$ is reached in $\dctmc$.

\subsection{Constructing \dCTMC}
%

As the first step in this alternative direction, we provide two basic techniques that significantly decrease the error for interval distributions (compared to standard PH algorithms that are by definition IPH algorithms as well).
%
%
The first technique deals with interval distributions bounded from below.

\begin{figure}[tb]
  \centering
\iffig \input{figures/figure3.pgf} \fi
\caption{On the left, there is an IPH approximation of the event \textsf{ack} using the algorithm \algonep{PhFit} with 3 phases. The discrete event $d$ has delay $4.05$. On the right, there is the whole \dCTMC{} with discrete-time events \textsf{timeout} and \textsf{d}. The model is obtained similarly as the CTMC in Figure~\ref{fig:example-gsmp}.}
\label{fig:example-dctmc}
\end{figure}
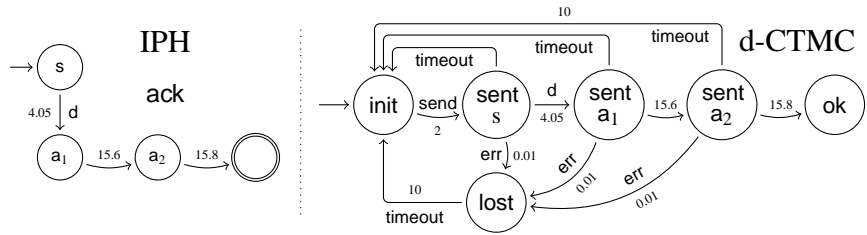

\begin{figure}[tb]
  \centering
\iffig 
\input{figures/figure4.pgf} 
\fi
\caption{The comparison of the approximations of the event \textsf{ack} using the algorithms PhFit and \algonep{PhFit} with 30 phases. On the left, there is the density of the original distribution as well as the both approximated densities. 
In the centre, there is for both approximations the difference of the original and the approximate cumulative distribution function. 
Notice that in point $x$ the plot displays for each algorithm the error we obtain when measuring the probability that the event occurs within time $x$ (rising as high as $0.5$).
%
%
On the right, there is for both approximations the difference of the original and the approximate density. The integral of this curve is the absolute density difference (\add) that we study.}
\label{fig:method-shifted}
\end{figure}
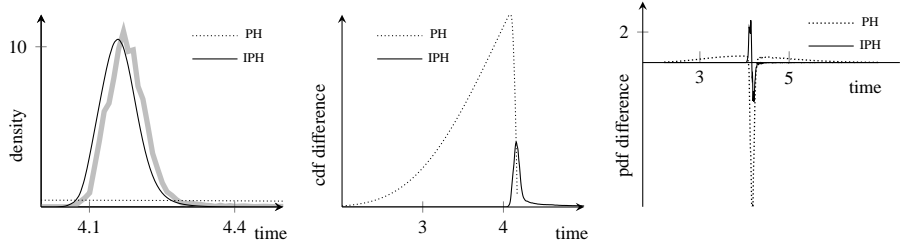

\begin{figure}[tb]
  \centering
\iffig 
\input{figures/figure5.pgf} 
\fi
\caption{The comparison of the approximations of the distribution uniform on $[0,2]$ using the algorithms PhFit and \algtwop{PhFit}. It goes along the same lines as in Figure~\ref{fig:method-shifted}.}
\label{fig:method-sliced}
\end{figure}
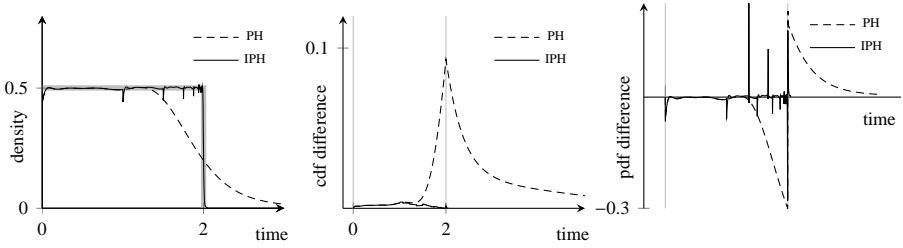

\begin{figure}[tb]
  \centering
\input{figures/figure6.pgf} 
\caption{On the left, the uniform distribution on $[0,2]$ is sliced into three subintervals. With the solid line, there is the whole density and its PH approximation corresponding to the CTMC below. With the dashed and dotted line, there are the conditional densities given the event does not occur before $1$ and $1.5$, respectively, and their PH approximations. Their corresponding CTMC are the same as the CTMC below, only with rates $2x$ and $4x$ larger, respectively.
This is clarified on the right, in the complete \dCTMC{} approximation with all 3 components (sharing the absorbing state $0$).}
\label{fig:example-uniform}
\end{figure}
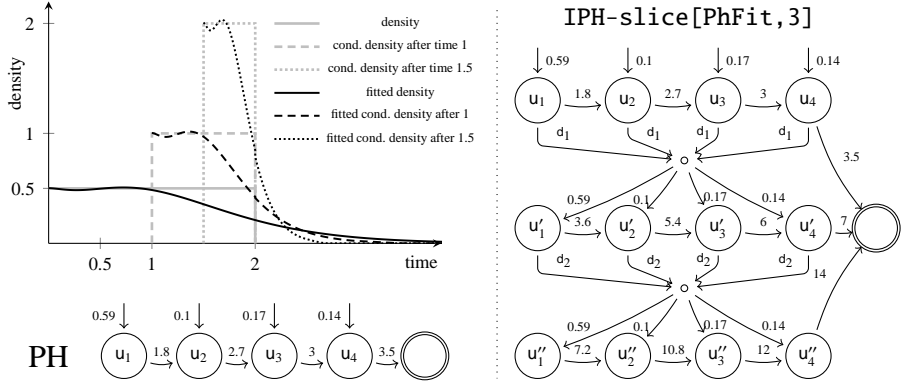

\subsubsection{Delay bounded from below}

For an event that cannot occur before some $l > 0$ and for a given number of phases $n > 1$, our algorithm works as follows.
Let $\ctmc = (\states,\events,\sched,\occur,\distribution_0)$ be a chain with $n-1$ phases fitted by some other tool \texttt{FIT} to the density on the interval $[l,\infty)$. We output a {\dCTMC} $(\states \uplus \{s_0\},\events\uplus \{d\},\sched',\occur',\distribution'_0)$ with $n$ states that starts with probability $\distribution'_0(s_0) = 1$ in the newly added state $s_0$ in which only the newly added event $d$ is scheduled, i.e. $\sched'(s_0) = \{d\}$; the event $d$ has delay $\delay(d) = l$ and after it occurs, the chain moves according to the initial distribution of $\ctmc$, i.e. $\occur'(s_0,d) = \distribution_0$; $\sched'$ and $\occur'$ coincide with $\sched$ and $\occur$ elsewhere.
%
A pseudo-code for this algorithm \algonep{FIT} is given in Appendix~\ref{app:alg}.

\paragraph{Example (continued)} To obtain the \dCTMC{} approximation of the GSMP model of the Alternating bit protocol, we only need to approximate the event \textsf{ack} since \textsf{timeout} is a discrete-time event. To show an example of the technique, the approximation of the event \textsf{ack} using the algorithm \algonep{PhFit} as well as the whole resulting \dCTMC{} is depicted in Figure~\ref{fig:example-dctmc}.
Since \algone is using the phase-type approximation only on the ``simple'' part of the density function, it gets much better results.
For instance for $30$ phases it yields approx. 4x smaller error compared to the best results of PH algorithms. In Figure~\ref{fig:method-shifted} we provide a more detailed comparison.


\subsubsection{Delay bounded from above}

For an event that cannot occur after some $u < \infty$, our algorithm \algtwop{FIT,p} slices the interval $[0,u]$ using discrete-time events into $p$ subintervals $[0,\frac{1}{2}u]$, $[\frac{1}{2}u,\frac{3}{4}u]$, $[\frac{3}{4}u,\frac{7}{8}u], \ldots, [(1-\frac{1}{2}^{p-2})u, (1-\frac{1}{2}^{p-1})u]$, $[(1-\frac{1}{2}^{p-1})u, u]$. Their length decreases exponentially with the last two subintervals having the same length. Corresponding to these intervals, we build a sequence of components $\ctmc_1, \ldots, \ctmc_p$ that is traversed by a sequence of discrete-time events $d_1,\ldots,d_{p-1}$ as the time flows. The component of each subinterval $[a,b]$ has $n/p$ phases and is fitted by \texttt{FIT} to the \emph{conditional} density of the \emph{remaining} delay given the event has not occurred on $[0,a)$. 
%
Consider the example from Figure~\ref{fig:example-uniform}.
The uniform distribution on $[0,2]$ has density $0.5$ in this interval and $0$ elsewhere. When already $1.5$ time units pass, the conditional density of the remaining delay equals $2$ on $[0,0.5]$ and $0$ elsewhere. 

This algorithm \algtwop{FIT,p} is formally described in Appendix~\ref{app:alg}. Example output of \algtwop{PhFit,3} on the above mentioned uniform distribution is depicted in Figure~\ref{fig:example-uniform}.
Similarly to the previous technique, it provides approximately 8x better results than the standard PH fitting as demonstrated in Figure~\ref{fig:method-sliced}.
Note that we can easily combine the two techniques for distributions bounded both from below and above such as uniform on $[5,6]$. It suffices to apply \algonep{\algtwop{FIT,slices}}.

Let us provide two remarks on this technique. 
First, notice that a standard fitting tool is applied on the conditional densities. However, a standard fitting tool tries to minimize the error also \emph{beyond} the subinterval we are dealing with which may lead to suboptimal approximation \emph{on} the subinterval. Modification of a PH algorithm addressing this issue might decrease the error of \algtwo even more.
Second, dividing the support of the distribution into subintervals of exponentially decreasing length is a heuristic that works well for distributions where the density does not vary much. For substantial discontinuities in the density, one should consider dividing the support in the points of discontinuity.
%
%
%
%
%
%
%
Next, we briefly review the analysis methods for \dCTMC.


%% file: figures/figure3.pgf
\begin{tikzpicture}[font=\scriptsize]

\tikzstyle{state} = [draw, circle, minimum size=2.5em, outer sep=0.2em,font=\footnotesize, inner sep=0.2em]
\tikzstyle{smallstate} = [state, minimum size=1.9em,font=\scriptsize, inner sep=0.2em]
\tikzstyle{trans} = [->]
\tikzstyle{small} = [font=\tiny]
\tikzstyle{multi} = [text width=15,text centered, font=\scriptsize]
\tikzstyle{fakemulti} = [minimum size=2.5em]

\begin{scope}[shift={(0.2,0.5)}] 
		\node[font=\large] at (3.5,-1.15) {IPH};
		\node[font=\normalsize] at (3.5,-1.85) {\ack};
	\begin{scope}[shift={(0,-1.5)}] 
		\node[state,smallstate] (start) at (2.1, 0) {\start};
		\node[state,smallstate] (aone) at (2.1, -1.2) {\aone};
		\node[state,smallstate] (atwo) at (3.4,-1.2) {\atwo};
		\node[state,smallstate,accepting,outer sep=0.2em] (athree) at (4.7,-1.2) {};

		\path  
			(start) edge[trans] node[auto] {$\discrete$} node[auto,swap,small] {$4.05$} (aone)
			(aone) edge[trans,bend right=15] node[auto,small] {$15.6$} (atwo)
			(atwo) edge[trans,bend right=15] node[auto,small] {$15.8$} (athree)
		;
		\draw [trans] ($(start.west)+(-0.3,0)$) -- (start);
	\end{scope}

\end{scope}

\begin{scope}[shift={(5.5,0)}] 
	\draw [dotted] (0,-0.5) -- (0,-3);
\end{scope}

\begin{scope}[shift={(6.6,0)}] 

	\node[font=\large] at (5.5,-0.65) {d-CTMC};
	
	\begin{scope}[shift={(0,-1.5)}] 
		\node[state, fakemulti] (init) at (0,-0) {\init};
		\node[state, multi] (sent1) at (1.5,0) {\footnotesize{\sent} \\ s};
		\node[state, multi] (sent2) at (3,0) {\footnotesize{\sent} \\ \aone};
		\node[state, multi] (sent3) at (4.5,0) {\footnotesize{\sent} \\ \atwo};
		\node[state, fakemulti] (lost) at (1.5,-1.3) {\lost};
		\node[state, fakemulti] (ok) at (6,0) {\ok};
		
		\path  
			(init) edge[trans,bend right=15] node[auto] {\send} node[auto,swap,small] {2} (sent1)
			(sent1) edge[trans] node[auto] {$\discrete$} node[auto,swap,small] {$4.05$} (sent2)
			(sent2) edge[trans,bend right=15] node[auto,small] {$15.6$} (sent3)
			(sent3) edge[trans,bend right=15] node[auto,small] {$15.8$} (ok)
 			(sent1) edge[trans,bend left=15] node[left=-1] {\err} node[right=-1,small] {$0.01$} (lost)
 			(sent2) edge[trans,bend left=30] node[above,sloped,pos=0.4] {\err} node[below,sloped,small,pos=0.4] {$0.01$} (lost)
 			(sent3) edge[trans,bend left=30] node[above,sloped,pos=0.4] {\err} node[below,sloped,small,pos=0.4] {$0.01$} (lost)
		;
 		\draw [rounded corners=2,trans] (lost) -| node[auto,below,pos=0.3] {\timeout} node[auto,above,pos=0.3,small] {10} (init);
		\draw [rounded corners=2,trans] (sent1.north) -- ($(sent1.north)+(-0,0.25)$) -| node[auto,below=-1,pos=0.25] {\timeout}  (init.75);
		\draw [rounded corners=2,trans] (sent2.north) -- +(-0,0.4) -| node[auto,below=-1,pos=0.1] {\timeout} node[auto,above=4,pos=0.1,small] {10} (init.90);
		\draw [rounded corners=2,trans] (sent3.north) -- +(-0,0.55) -| node[auto,below=-1,pos=0.06] {\timeout}  (init.105);
 		\draw [trans] ($(init.west)+(-0.4,0)$) -- (init);

	\end{scope}
\end{scope}

\end{tikzpicture}

%% file: figures/figure4.pgf
\begin{tikzpicture}[font=\scriptsize]

\begin{scope}[shift={(-4,0)}] 

\begin{axis}[
	restrict x to domain=4:4.5,
	ymin=0,ymax=12,
	no markers,
	axis y line=left,axis x line=bottom,
 	xtick=\empty, ytick=\empty,
	width=4.8cm,
]

\addplot[line width=2pt, lightgray] plot file {\figpath/tikz_data.pdf.txt};
\end{axis}

\begin{axis}[
	restrict x to domain=4:4.5,
	ymin=0,ymax=12,
	no markers,
	axis y line=left,axis x line=bottom,
 	ytick={10}, xtick={4.1, 4.4},
	width=4.8cm,
 	xlabel=time, ylabel=density,
 	xlabel style={shift={(0.45,0.15)}},ylabel style={shift={(-0.1,-0.36)}},
	legend style={font=\tiny,draw=none,fill=none}
]

\addplot[black,smooth, densely dotted] file {\figpath/tikz_ph.pdf.txt};
\addplot[black,smooth] file {\figpath/tikz_iph.pdf.txt};

\addlegendentry{PH}
\addlegendentry{IPH}

\end{axis}
\end{scope}

\begin{scope}[shift={(0,0)}] 

\begin{axis}[
	restrict x to domain=2:5,
	restrict y to domain=0:0.55,
	no markers,
	axis y line=left,axis x line=bottom,
 	xtick={3,4},ytick={0.5},
	width=4.8cm,
 	xlabel=time, ylabel=cdf difference,
 	xlabel style={shift={(0.45,0.15)}},ylabel style={shift={(-0.1,-0.36)}},
	legend style={font=\tiny,draw=none,fill=none, at={(0.5,0.98)}}
]

\addplot[black,smooth, densely dotted] file {\figpath/tikz_ph.cdf.diffs.txt};
\addplot[black,smooth] file {\figpath/tikz_iph.cdf.diffs.txt};

\addlegendentry{PH}
\addlegendentry{IPH}

\end{axis}
\end{scope}

\begin{scope}[shift={(4,0)}] 
\begin{axis}[
	restrict x to domain=2.2:7,
	restrict y to domain=-10.5:4,
	ymax=4,
	no markers,
	axis y line=left,axis x line*=center,
 	xtick={3,5},ytick={-10,2},
	width=5cm,
 	xlabel=time, ylabel=pdf difference,
 	xlabel style={shift={(0.45,0.85)}},ylabel style={shift={(0,-0.36)}},
	legend style={font=\tiny,draw=none,fill=none}
]

\addplot[black,smooth, densely dotted,line width=0.5pt] file {\figpath/tikz_ph.pdf.diffs.txt};
\addplot[black,smooth] file {\figpath/tikz_iph.pdf.diffs.txt};

\addlegendentry{PH}
\addlegendentry{IPH}

\end{axis}
\end{scope}

\end{tikzpicture}

%% file: figures/figure5.pgf
\begin{tikzpicture}[font=\scriptsize]

\begin{scope}[shift={(-4,0)}] 

\begin{axis}[
	ymin=0,ymax=0.8,xmin=0, xmax=3,
	no markers,
	axis y line=left,axis x line=bottom,
 	xtick=\empty, ytick=\empty,
	width=4.8cm,
]

\addplot[line width=2pt, lightgray] plot coordinates {(0,0.5) (2,0.5) (2,0)};
\end{axis}

\begin{axis}[
	ymin=0,ymax=0.8,xmin=0, xmax=3,
	no markers,
	axis y line=left,axis x line=bottom,
	extra x ticks={4.05},
	extra x tick style={grid=major},
	extra x tick labels={$l$},
 	xtick={0,2}, ytick={0,0.5},
	width=4.8cm,
 	xlabel=time, ylabel=density,
 	xlabel style={shift={(0.45,0.15)}},ylabel style={shift={(-0.1,-0.36)}},
	legend style={font=\tiny,draw=none,fill=none}
]

\addplot[black,smooth, densely dashed] file {\figpath/fig5_pdf.data};
\addplot[black] file {\figpath/fig5_pdf_i.data};

\addlegendentry{PH}
\addlegendentry{IPH}

\end{axis}
\end{scope}

\begin{scope}[shift={(0,0)}] 

\begin{axis}[
	ymin=0,ymax=0.12,xmin=-0.22, xmax=5,
	no markers,
	axis y line=left,axis x line=bottom,
 	xtick={0.5,1,2,4},ytick={0.5,1,2},
	extra x ticks={0,2},
	extra x tick style={grid=major},
	extra x tick labels={0,2},
 	xtick=\empty, ytick={0.1},
	width=4.8cm,
 	xlabel=time, ylabel=cdf difference,
 	xlabel style={shift={(0.45,0.15)}},ylabel style={shift={(-0.1,-0.36)}},
	legend style={font=\tiny,draw=none,fill=none, at={(0.98,0.98)}}
]

 \addplot[black,smooth, densely dashed] file {\figpath/fig5_cdf_diff.data};
 \addplot[black,smooth] file {\figpath/fig5_cdf_diff_i.data};

\addlegendentry{PH}
\addlegendentry{IPH}

\end{axis}
\end{scope}

\begin{scope}[shift={(4,0)}] 
\begin{axis}[
	restrict y to domain=-0.32:0.32,
	restrict x to domain=-0.2:3.5,
	no markers,
	axis y line=left,axis x line*=center,
	extra x ticks={0,2},
	extra x tick style={grid=major,shift={(0.1,0)}},
 	extra x tick labels={,},
 	xtick=\empty, ytick={0.3,-0.3},
	width=5cm,
 	xlabel=time, ylabel=pdf difference,
 	xlabel style={shift={(0.5,0.7)}},ylabel style={shift={(0,-0.36)}},
	legend style={font=\tiny,draw=none,fill=none}
]

 \addplot[black,smooth, densely dashed] file {\figpath/fig5_pdf_diff.data};
 \addplot[black,smooth] file {\figpath/fig5_pdf_diff_i.data};

\addlegendentry{PH}
\addlegendentry{IPH}

\end{axis}
\end{scope}

\end{tikzpicture}

%% file: figures/figure6.pgf
\begin{tikzpicture}[font=\scriptsize]

\tikzstyle{state} = [draw, circle, minimum size=2.5em, outer sep=0.2em,font=\footnotesize, inner sep=0.2em]
\tikzstyle{smallstate} = [state, minimum size=1.9em,font=\scriptsize, inner sep=0.2em]
\tikzstyle{trans} = [->]
\tikzstyle{small} = [font=\tiny]
\tikzstyle{multi} = [text width=15,text centered, font=\scriptsize]
\tikzstyle{fakemulti} = [minimum size=2.5em]

\begin{scope}[shift={(-0.8,-3.4)}] 
\begin{axis}[
	ymin=0,ymax=2.2,xmin=0, xmax=3.8,
	no markers,
	axis y line=left,axis x line=bottom,
	xtick={0.5,1,2,4},ytick={0.5,1,2},
	width=6.8cm, height=4.8cm,
	xlabel=time, ylabel=density,
	xlabel style={shift={(0.45,0.15)}},ylabel style={shift={(0.10,-0.25)}},
	legend style={font=\tiny,draw=none,at={(1.12,0.98)}}
]

\addplot[line width=1pt, lightgray] plot coordinates {(0,0.5) (2,0.5) (2,0)};
\addplot[line width=1pt, lightgray,densely dashed] plot coordinates {(1,0) (1,1) (2,1) (2,0)};
\addplot[line width=1pt, lightgray,densely dotted] plot coordinates {(1.5,0) (1.5,2) (2,2) (2,0)};

\addplot[line width=0.75pt,black] table[x=x,y=four] {\figpath/figure6.data};
\addplot[line width=0.75pt,black,densely dashed] table[x=x,y=four2] {\figpath/figure6.data};
\addplot[line width=0.75pt,black,densely dotted] table[x=x,y=four3] {\figpath/figure6.data};

\addlegendentry{density}
\addlegendentry{cond. density after time $1$}
\addlegendentry{cond. density after time $1.5$}
\addlegendentry{fitted density}
\addlegendentry{fitted cond. density after $1$}
\addlegendentry{fitted cond. density after $1.5$}

\end{axis}
\end{scope}

\begin{scope}[shift={(0.2,-3.4)}] 
		\node[font=\large] at (-1,-1.5) {PH};
	\begin{scope}[shift={(0,-1.5)}] 
		\node[state,smallstate] (uone) at (0, 0) {$\uone$};
		\node[state,smallstate] (utwo) at (1, 0) {$\utwo$};
		\node[state,smallstate] (uthree) at (2,0) {$\uthree$};
		\node[state,smallstate] (ufour) at (3,0) {$\ufour$};
		\node[state,smallstate,accepting,outer sep=0.2em] (ufive) at (4,0) {};

		\path  
			(uone) edge[trans,bend right=15] node[auto,small] {$1.8$} (utwo)
			(utwo) edge[trans,bend right=15] node[auto,small] {$2.7$} (uthree)
			(uthree) edge[trans,bend right=15] node[auto,small] {$3$} (ufour)
			(ufour) edge[trans,bend right=15] node[auto,small] {$3.5$} (ufive)
		;
   		\draw [<-]  (uone) --  node[auto,small] {$0.59$} +(0,0.7);
   		\draw [<-]  (utwo) --  node[auto,small] {$0.1$} +(0,0.7);
   		\draw [<-]  (uthree) --  node[auto,small] {$0.17$}+(0,0.7);
   		\draw [<-]  (ufour) --  node[auto,small] {$0.14$}+(0,0.7);
	\end{scope}

\end{scope}

\begin{scope}[shift={(5.16,0)}] 
	\draw [dotted] (0,-0.3) -- (0,-5.2);
\end{scope}

\begin{scope}[shift={(5.7,0)}] 

	\node[font=\normalsize] at (2,-0.4) {\texttt{IPH-slice[PhFit,3]}};
	
	\begin{scope}[shift={(0,-1.5)}] 
		\node[smallstate] (u11) at (0,-0) {$\uone$};
		\node[smallstate] (u12) at (1.2,0) {$\utwo$};
		\node[smallstate] (u13) at (2.4,0) {$\uthree$};
		\node[smallstate] (u14) at (3.6,0) {$\ufour$};

		\node[smallstate, accepting] (ok) at (4.5,-1.7) {};
		
		\path  
			(u11) edge[trans,bend right=10] node[auto,small] {$1.8$} (u12)
			(u12) edge[trans,bend right=10] node[auto,small] {$2.7$} (u13)
			(u13) edge[trans,bend right=10] node[auto,small] {$3$} (u14)
			(u14) edge[trans,bend right=10] node[auto,small] {$3.5$} (ok)
		;
   		\draw [<-]  (u11) --  node[auto,small,swap] {$0.59$} +(0,0.7);
   		\draw [<-]  (u12) --  node[auto,small,swap] {$0.1$} +(0,0.7);
   		\draw [<-]  (u13) --  node[auto,small,swap] {$0.17$}+(0,0.7);
   		\draw [<-]  (u14) --  node[auto,small,swap] {$0.14$}+(0,0.7);
	\end{scope}

	\begin{scope}[shift={(0,-3.2)}] 
		\node[smallstate] (u21) at (0,-0) {$\uone'$};
		\node[smallstate] (u22) at (1.2,0) {$\utwo'$};
		\node[smallstate] (u23) at (2.4,0) {$\uthree'$};
		\node[smallstate] (u24) at (3.6,0) {$\ufour'$};

		\node[draw,inner sep=0.1em,circle,outer sep=0.4em] (dot2) at (1.95,0.9) {};
		
		\path  
			(u21) edge[trans,bend right=10] node[auto,small] {$3.6$} (u22)
			(u22) edge[trans,bend right=10] node[auto,small] {$5.4$} (u23)
			(u23) edge[trans,bend right=10] node[auto,small] {$6$} (u24)
			(u24) edge[trans,bend right=10] node[auto,small] {$7$} (ok)

			(dot2)  edge[trans,bend left=5] node[above left=-2,small,pos=0.75] {$0.59$} (u21)
			(dot2)  edge[trans,bend left=5] node[left=-1,small,pos=0.75] {$0.1$} (u22)
			(dot2)  edge[trans,bend right=5] node[right=-1.5,small,pos=0.75] {$0.17$} (u23)
			(dot2)  edge[trans,bend right=5] node[above right=-2,small,pos=0.75] {$0.14$} (u24)
		;

  		\draw [rounded corners=2,trans] (u11) -- +(0,-0.6) -- node[small,auto,above,pos=0.2] {$\discrete_1$}  (dot2);
  		\draw [rounded corners=2,trans] (u12) -- +(0,-0.5) -- node[small,auto,above,pos=0.6] {$\discrete_1$}  (dot2);
  		\draw [rounded corners=2,trans] (u13) -- +(0,-0.5) -- node[small,auto,above,pos=0.6] {$\discrete_1$}  (dot2);
  		\draw [rounded corners=2,trans] (u14) -- +(0,-0.6) -- node[small,auto,above,pos=0.2] {$\discrete_1$}  (dot2);

	\end{scope}

\begin{scope}[shift={(0,-4.9)}] 
		\node[smallstate] (u31) at (0,-0) {$\uone''$};
		\node[smallstate] (u32) at (1.2,0) {$\utwo''$};
		\node[smallstate] (u33) at (2.4,0) {$\uthree''$};
		\node[smallstate] (u34) at (3.6,0) {$\ufour''$};

		\node[draw,inner sep=0.1em,circle,outer sep=0.4em] (dot3) at (1.95,0.9) {};
		
		\path  
			(u31) edge[trans,bend right=10] node[auto,small] {$7.2$} (u32)
			(u32) edge[trans,bend right=10] node[auto,small] {$10.8$} (u33)
			(u33) edge[trans,bend right=10] node[auto,small] {$12$} (u34)
			(u34) edge[trans,bend left=10] node[auto,small] {$14$} (ok)

			(dot3)  edge[trans,bend left=5] node[above left=-2,small,pos=0.75] {$0.59$} (u31)
			(dot3)  edge[trans,bend left=5] node[left=-1,small,pos=0.75] {$0.1$} (u32)
			(dot3)  edge[trans,bend right=5] node[right=-1.5,small,pos=0.75] {$0.17$} (u33)
			(dot3)  edge[trans,bend right=5] node[above right=-2,small,pos=0.75] {$0.14$} (u34)
		;

  		\draw [rounded corners=2,trans] (u21) -- +(0,-0.6) -- node[small,auto,above,pos=0.2] {$\discrete_2$}  (dot3);
  		\draw [rounded corners=2,trans] (u22) -- +(0,-0.5) -- node[small,auto,above,pos=0.6] {$\discrete_2$}  (dot3);
  		\draw [rounded corners=2,trans] (u23) -- +(0,-0.5) -- node[small,auto,above,pos=0.6] {$\discrete_2$}  (dot3);
  		\draw [rounded corners=2,trans] (u24) -- +(0,-0.6) -- node[small,auto,above,pos=0.2] {$\discrete_2$}  (dot3);

	\end{scope}

\end{scope}

\end{tikzpicture}

%% file: methods.tex
\subsection{Analysing \dCTMC}
\label{sec-methods}

The existing theory and algorithms applicable to analysis of \dCTMC{} are a crucial part of our alternative IPH approximation method. Extending the knowledge in this direction is out of scope of this paper, here we only summarize the state-of-the-art of transient and stationary analysis.


The method of \emph{supplementary variables}~\cite{cox1955supplementary,GL:dspn-analysis-supplementary-variables,lindemann1999transient} analyses the continuous state-space $\states\times(\Rsetpo)^\events$ extended by the remaining times $\val(e)$ until each currently active discrete-time event $e$ occurs. The system is described by partial differential equations and solved by discretization in the tool DSPN\-Express~2.0~\cite{LRT:dspnexpress-tool}. A more elaborate method of \emph{stochastic state classes}~\cite{sassoli2007close,alur2006bounded,horvath2010transient,horvath2012transient} implemented in the tool Oris~\cite{bucci2010oris} studies the continuous state-space model at moments when events occur (defining an embedded Markov chain). In each such moment, multidimensional densities over $\val(e)$ are symbolically derived. The embedded chain is finite iff the system is regenerative, approximation is applied otherwise.


If the \dCTMC{} has at most one discrete-time event active at a time (e.g. when only one event is approximated by IPH), one can apply the efficient method of \emph{subordinated Markov chains}~\cite{marsan1987petri}.
 It builds the embedded Markov chain using transient analysis of CTMC, similarly to the analysis of CTMC observed by a one-clock timed automaton~\cite{chen2009quantitative}.
In the tool Sabre~\cite{guet2012delayed}, this method is extended to parallel discrete-time events by approximating them 
using one discrete-time event $\Delta$~\cite{haddad2005performance} that is active in all states and emulates other discrete-time events. An event $e$ occurs with the $\lfloor \delay(e)/\delay(\Delta) \rfloor$-th occurrence of $\Delta$ after initialization of $e$. Note that this corresponds to discretizing time for the discrete-time events while leaving the exponential events intact. 
%

As some of the methods are recent, no good comparison of these methods exists. Based on our preliminary experiments, we apply in Section~\ref{sec-results} the tool Sabre.

%% file: results.tex
\section{Experimental Evaluation}
\label{sec-results}

In this section we 
evaluate the reduction of the state space and hence the reduction of the time needed for the analysis
when using IPH compared to PH. Precisely, (1) we inspect the growth of the
state space of both IPH and PH approximations when decreasing the tolerated
error; (2) for a fixed tolerated error, we examine the growth of the state
space of the PH approximation when increasing the shift of a shifted
distribution; and (3) for a fixed model and a fixed PCTL property we compare
the running time of the analysis of \dCTMC{} yielded by IPH and the running time of the analysis of CTMC yielded by PH when increasing the number of phases.

We consider the distributions from the previous sections, namely
the shifted distribution of the event \textsf{ack} addressed by the \algone
algorithm and the distribution uniform on $[0,2]$ addressed by the \algtwo
algorithm.  The uniform distribution is specified simply by its formula
whereas the density of the event \textsf{ack} is based on real data. Using the
Unix \texttt{ping} command, we collected 10000 successful ICMP response
times of a web server (www.seznam.cz, the most visited web portal in the Czech Republic). The data set has mean $4.19$~ms, standard deviation
$0.314$~ms, variance $0.0986$, coefficient of variation $0.075$~ms, and the
shortest time is $4.06$~ms (see Figure~\ref{fig:method-shifted}).


\setlength{\tabcolsep}{6pt}
\begin{table}[tb]
\centering
\begin{tabular}{lccc}
\toprule
PH fitting tool & (\add) for event \textsf{ack} & (\add) for uniform distribution & CPU time \\
\noalign{\vspace{2pt}}
\hline
\noalign{\vspace{2pt}}
EMpht & 1.7957 &  1.8980 & over one day \\
G-FIT & 1.6100 & 0.1603 & 4 min 49 s \\ 
momfit & 1.8980 & 0.5820 & 1 day\\
PhFit & 1.6518 & 0.1868 & 4.33 s\\
\bottomrule
\vspace{0.00pt}
\end{tabular}
\caption{The (\add) errors and CPU time for different PH tools fitting by 30 phases. 
}
\label{tab:results}
\end{table}

To get reliable results, we need to compare IPH with state-of-the-art tools for 
continuous PH fitting. For our experiments, we considered the tools
\emph{EMpht}~\cite{asmussen1996fitting},
\emph{G-FIT}~\cite{thummler2006gfit},
\emph{momfit}~\cite{horvath2007momfit}, and
\emph{PhFit}~\cite{horvath2002phfit}.
We ran the tools to produce PH approximations of the two events with 30
phases (we chose such a small number of phases because for some tools it
already took a substantial amount of time).  Based on the results shown in
Table~\ref{tab:results}, we have selected PhFit as the baseline tool.  Most
of the tools achieve similar precision, however PhFit significantly
outperforms all others regarding the CPU time\footnote{The analysis has been
  performed on Red Hat Enterprise Linux 6.5 running on a server with 8
  processors Intel Xeon X7560 2.26GHz (each with 8 cores) and shared 448 GiB
  DDR3 RAM.}.  

\subsection{Growth of the state space when decreasing error}

\begin{figure}[tb]
  \centering
\iffig \input{figures/figure7.pgf} \fi
\caption{The (logarithmically scaled) relationship between the size of the state space and the
  error obtained. 
  For the uniform distribution, we show the results for different numbers of
  slices, each with the same number of phases. The plotted number of phases
  is the sum of the phases within all used slices. The error for the optimal
  number of slices is plotted in bold.}
\label{fig:results-error-vs-phases}
\end{figure}
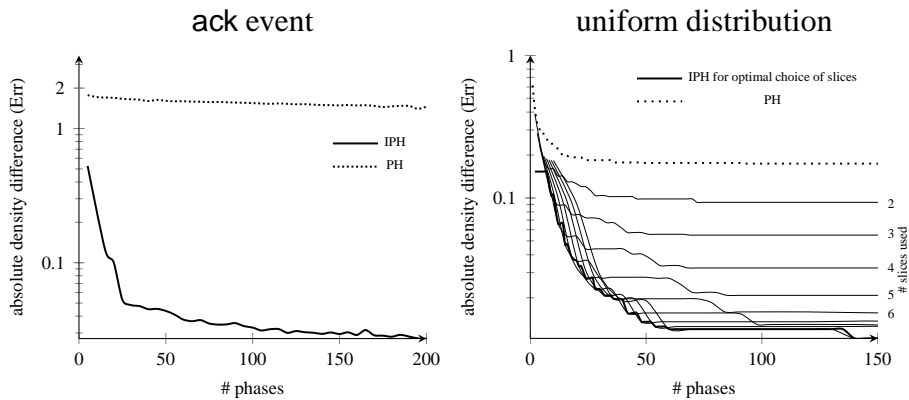

In the first experiment, we focus on the size of the state space necessary
to fit the distributions up to a decreasing error.
%
The decreasing errors (\add) when increasing the number of phases, i.e. the state space, are shown in Figure~\ref{fig:results-error-vs-phases}.
Both our IPH algorithms exhibit a fast decrease of the error (note that the
scales are logarithmic). Observe that the continuous PH method does not
perform particularly well on the event \textsf{ack} obtained as a
real-world example since the absolute density difference of two densities can never exceed $2$. For the uniform distribution, we show the
results for different numbers of slices used in the \algtwo algorithm.  
According to our
experiments on the uniform distribution, a finer slicing with less phases in
each slice is better than a coarser one with more phases in each slice,
whenever each slice is fitted by at least $4$ phases.

\subsection{Growth of the state space when increasing the shift}

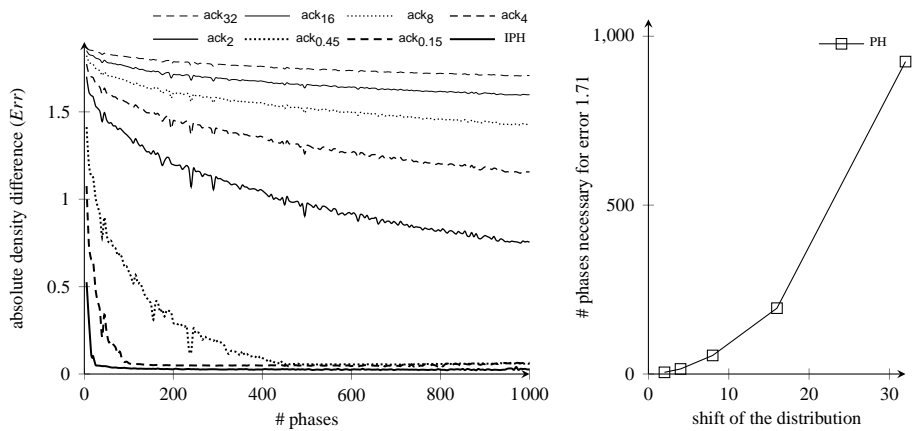
\begin{figure}[tb]
  \centering
\iffig \input{figures/figure8.pgf} \fi
\caption{The growth of the state space when increasing the shift.  On the
  left, there is the dependence of the error on the size of the state space
  for distributions with different shifts. Note that the IPH fitting does not
  depend on the shift. On the right, there is the growth of the state space
  when increasing the shift and fixing the PH fitting error to $1.71$.}
\label{fig:results-shift-vs-phases}
\end{figure}

In the second experiment, we analyse the growth of the state space when
increasing the shift of a shifted distribution. In other words, \emph{how
  much larger model we get when we try to fit with a fixed error an event
  with lower coefficient of variation?} We took the distribution of the
\textsf{ack} event and shifted the data to obtain a sequence of events
$\mathsf{ack_{0.15}}, 
\cdots , \mathsf{ack_{4}}, \cdots \mathsf{ack_{32}}$ where $\mathsf{ack_i}$ has zero density on the
interval $[0,i]$.  Note that compared to \textsf{ack} we shifted the data in
both directions as $\mathsf{ack} \approx \mathsf{ack_4}$.  The results in
Figure~\ref{fig:results-shift-vs-phases} confirm a quadratic relationship
between the shift and the necessary number of phases for the PH
approximation~\cite{guet2012delayed}. 

The quadratic relationship can be supported by the following
explanation. Assume we want to approximate a discrete distribution with shift
$s$ by a PH distribution. Due to \cite{AS87Erlang}, the best PH distribution
for this purpose is the Erlang distribution, the chain of $k$ phases with
exit rates $k/s$. Since (\add) does not work in this setting (density is
not defined for discrete distributions), we use another common metric -
matching moments. Here the goal is to exactly match the mean and minimize the
difference of variance. Since the variance of the discrete distribution is zero,
the error for $k$ phases is the variance of the Erlang distribution,
i.e. $s^2/k$. To get the same error for a discrete distribution with $n$-times increased shift $n \cdot s$, we need $n^2 \cdot k$ phases as $(n\cdot s)^2/(n^2\cdot k)=s^2/k$.

\subsection{Time requirements and error convergence when increasing state space}

\setlength{\tabcolsep}{6pt}
\begin{table}[tb]
\centering
\begin{tabular}{cccccc}
\toprule
\multicolumn{3}{c}{PRISM on CTMC} & \multicolumn{3}{c}{Sabre on \dCTMC{}} \\
phases  & result & CPU time & phases & result & CPU time \\
\noalign{\vspace{2pt}}
\hline
\noalign{\vspace{2pt}}
100 & 0.527 & 6.37 s & 5 & 0.695 & 41s \\
200 & 0.541 & 14.92 s & 10 & 0.730 &  1 min 24 s \\ 
500 & 0.562 & 47.74 s & 20 & 0.745 & 3 min 37 s \\
1000 & 0.585 & 2 min 55 s  & 30 & 0.758 & 10 min 30 s \\
2000 & 0.629 & 9 min 20s \\
3000 & 0.680 & 50 min 49s \\
5030 & 0.705 & 3h 14 min \\
10030 & 0.731 & 32 h 2 min  \\

\bottomrule
\vspace{0.00pt}
\end{tabular}
\caption{Probability of collision computed by unbounded reachability in CTMC derived
  using PH and in \dCTMC{} derived using IPH. The exact probability of collision is $0.7753$.}
\label{tab:prism_vs_sabre}
\end{table}

So far, we studied how succinct the IPH approximations are compared to PH.
%
One can
naturally dispute the impact of IPH approximation by saying that the
complexity of \dCTMC{} analysis 
is higher that the complexity of 
CTMC analysis. 
Here, we show an example where IPH in fact leads to a \emph{lower} overall analysis time.

We model two workstations competing for a shared channel. Each workstation
wants to transmit its data for which it needs $1.2$ seconds of an exclusive
use of the channel. Each workstation starts the transmission at a random
time. If one workstations starts its transmission when the other is
transmitting, a collision occurs.  Our goal is to compute the probability
of collision. 
%
For the transmission initiations, we again used the \texttt{ping} command (for two different servers) and obtained two distributions with zero density in the first $4.1$ seconds and the first $5.51$
seconds, respectively.


We approximated the model using both PH and IPH and subsequently run
analysis in the tools PRISM~\cite{KNP11} and Sabre~\cite{guet2012delayed} that are according to our knowledge the best
tools for analysing large CTMC and \dCTMC{} models, respectively \footnote{To eliminate the effects of the implementation, the CTMC analysis was run also in Sabre. However it was much slower than in PRISM, thus results er omitted here. For details see arxiv version of the paper on ...}. In the \dCTMC{} model for Sabre,
we used IPH approximation for both transmission initiating distributions and a
discrete-time event for the $1.2$ seconds of transmission. The probability of
collision was computed by reachability analysis. In the CTMC model for PRISM, we used PH approximations for both transmission initiating distributions. Furthermore, as PRISM does not support nesting of time bounded until operator into until operator, we again needed to transform the problem into (unbounded) reachability analysis by incorporating the $1.2$ seconds of transmission time in the model. We
approximated the time by Erlang distribution with $1000$ phases (using different
number of phases causes at most 1\% error in the result). 


The results of our experiments are shown in Table
\ref{tab:prism_vs_sabre}. The exact probability of collision is $0.7753$ as computed 
directly from the data sampled by \texttt{ping} using a 
LibreOffice spreadsheet. 
Due to some numerical errors in the version of Sabre that we used, we were not able to get a lower error than 2\% even when using more than 
30 phases.
Note that the results we were able to obtain from PRISM have a more
then twice as high error\footnote{We did our best to make CTMC analysis
as quick as possible, we used parameters \texttt{-s -gs -maxiters
1000000 -cuddmaxmem 18000000} and set \texttt{PRISM\_JAVAMAXMEM} to \texttt{200000m}.}.
Moreover, the immense analysis times shown in Table
\ref{tab:prism_vs_sabre} do not include the durations of PH approximations. The largest approximation we were able to obtain using PhFit was for 3000 phases as for 4000 phases it did not finish within 5 days.
%
For 5030 and 10030 phases we 
thus constructed the approximations 
by concatenating an Erlang approximation of the shift with the 30 phases PH approximation of the remaining part (as it was obtained during IPH).
%
%
%
%
Overall, the results indicate that for models that are sensitive to precise approximation of the distributions, the IPH approximation can lead to a significantly faster analysis compared to PH approximation.

%% file: figures/figure7.pgf
\begin{tikzpicture}[font=\scriptsize]

\pgfplotsset{%
    x tick label style={/pgf/number format/1000 sep=\,},
    log base 10 number format code/.code={%
        $\pgfmathparse{10^(#1)}\pgfmathprintnumber{\pgfmathresult}$%
    }%
  }  

\begin{scope}[shift={(0,0)}] 

\begin{axis}[
  	ymode=log,
	ymin=0,ymax=3.5,xmin=0, xmax=200,
	no markers,
	axis y line=left,axis x line=bottom,
  	extra y ticks={2},
  	extra y tick labels={{2}},
 	width=6.2cm,
	title={\large \textsf{ack} event},
  	xlabel={\# phases}, ylabel={absolute density difference (\add)},
 	xlabel style={yshift=2},ylabel style={yshift=-12},
 	legend style={font=\tiny,draw=none,fill=none,at={(0.98,0.75)}}
]

\addplot[line width=0.75pt,black,smooth] file {\figpath/tikz_phase-errorlin-4.05.txt};
\addplot[line width=0.75pt,black,smooth,densely dotted] file {\figpath/tikz_phase-errorlin.txt};


\addlegendentry{IPH}
\addlegendentry{PH}

\end{axis}
\end{scope}

\begin{scope}[shift={(6,0)}] 

\begin{axis}[
  	ymode=log,
	ymin=0,ymax=1,xmin=0, xmax=150,
	axis y line=left,axis x line=bottom,
  	extra y ticks={2},
  	extra y tick labels={{2}},
 	width=6.2cm,
	title={\large uniform distribution},
  	xlabel={\# phases}, ylabel={absolute density difference (\add)},
 	xlabel style={yshift=2},ylabel style={yshift=-12},
 	legend style={font=\tiny,draw=none,fill=none}
]

\addplot[line width=0.75pt,black,smooth] table[x=iph,y=eiph] {\figpath/fig7_uniform_pdf.data};
\addplot[line width=0.75pt,black,smooth,dotted] table[x=ph,y=eph] {\figpath/fig7_uniform_pdf.data};
\addplot[line width=0.1pt,black,smooth] table[x=p2,y=e2,domain=50:80] {\figpath/fig7_uniform_pdf.data};
\addplot[line width=0.1pt,black,smooth] table[x=p3,y=e3] {\figpath/fig7_uniform_pdf.data};
\addplot[line width=0.1pt,black,smooth] table[x=p4,y=e4] {\figpath/fig7_uniform_pdf.data};
\addplot[line width=0.1pt,black,smooth] table[x=p5,y=e5] {\figpath/fig7_uniform_pdf.data};
\addplot[line width=0.1pt,black,smooth] table[x=p6,y=e6] {\figpath/fig7_uniform_pdf.data};
\addplot[line width=0.1pt,black,smooth] table[x=p7,y=e7] {\figpath/fig7_uniform_pdf.data};
\addplot[line width=0.1pt,black,smooth] table[x=p8,y=e8] {\figpath/fig7_uniform_pdf.data};
\addplot[line width=0.1pt,black,smooth] table[x=p9,y=e9] {\figpath/fig7_uniform_pdf.data};
\addplot[line width=0.1pt,black,smooth] table[x=p10,y=e10,domain=50:150] {\figpath/fig7_uniform_pdf.data};


\addlegendentry{IPH for optimal choice of slices}
\addlegendentry{PH}

\end{axis}

\node [font=\tiny,rotate=90] at (4.95,1.05) {\# slices used};
\node [font=\tiny] at (4.8,1.8) {2};
\node [font=\tiny] at (4.8,1.4) {3};
\node [font=\tiny] at (4.8,0.95) {4};
\node [font=\tiny] at (4.8,0.58) {5};
\node [font=\tiny] at (4.8,0.33) {6};

\end{scope}


%
%
%
%
%
%
%
%
%
%
%
%

\end{tikzpicture}

%% file: figures/figure8.pgf
\begin{tikzpicture}[font=\scriptsize]

\pgfplotsset{%
    x tick label style={/pgf/number format/1000 sep=\,},
    log base 10 number format code/.code={%
        $\pgfmathparse{10^(#1)}\pgfmathprintnumber{\pgfmathresult}$%
    }%
  }  

\begin{scope}[shift={(0,0)}] 

\begin{scope}[shift={(-7.5,0)}] 

\begin{axis}[
	ymin=0,ymax=1.9,xmin=0, xmax=1000,
	no markers,
	axis y line=left,axis x line=bottom,
 	width=7.5cm,
	height=6cm,
  	xlabel={\# phases}, ylabel={absolute density difference $(\add)$},
  	xlabel style={shift={(0,0.03)}},ylabel style={shift={(0,-0.08)}},
 	legend style={font=\tiny,draw=none,fill=none,	at={(1.03,1.13)},legend columns=4},
]

\addplot[line width=0.25pt,black,smooth,densely dashed] file {\figpath/tikz_phase-errorlin+28.txt};
\addplot[line width=0.25pt,black,smooth] file {\figpath/tikz_phase-errorlin+12.txt};
\addplot[line width=0.5pt,black,smooth, densely dotted] file {\figpath/tikz_phase-errorlin+4.txt};
\addplot[line width=0.5pt,black,smooth,densely dashed] file {\figpath/tikz_phase-errorlin.txt};
\addplot[line width=0.5pt,black,smooth,] file {\figpath/tikz_phase-errorlin-2.txt};
\addplot[line width=0.75pt,black,smooth,densely dotted] file {\figpath/tikz_phase-errorlin-3.6.txt};
\addplot[line width=0.75pt,black,smooth,densely dashed] file {\figpath/tikz_phase-errorlin-3.9.txt};
\addplot[line width=0.75pt,black,smooth] file {\figpath/tikz_phase-errorlin-4.05.txt};


\addlegendentry{$\mathsf{ack_{32}}$}
\addlegendentry{$\mathsf{ack_{16}}$}
\addlegendentry{$\mathsf{ack_{8}}$}
\addlegendentry{$\mathsf{ack_{4}}$}
\addlegendentry{$\mathsf{ack_{2}}$}
\addlegendentry{$\mathsf{ack_{0.45}}$}
\addlegendentry{$\mathsf{ack_{0.15}}$}
\addlegendentry{IPH}

\end{axis}
\end{scope}

\begin{scope}[shift={(0,0)}] 

\begin{axis}[
	ymin=0,ymax=1050,xmin=0, xmax=32,
	axis y line=left,axis x line=bottom,
 	width=5cm,
	height=6.3cm,
  	xlabel={shift of the distribution}, ylabel={\# phases necessary for error 1.71},
  	xlabel style={shift={(0,0.03)}},ylabel style={shift={(0,-0.08)}},
 	legend style={font=\tiny,draw=none,fill=none}
]

\addplot[mark=square] coordinates {(2,5)(4,15)(8,55)(16,195)(32,925)};



\addlegendentry{PH}

\end{axis}
\end{scope}
\end{scope}


%
%
%
%
%
%
%
%
%
%
%
%

\end{tikzpicture}

%% file: app.tex
\section{Details on algorithms}
\label{app:alg}

Let us here give the pseudo-codes for algorithms \algone and \algtwo described in Section~\ref{sec-IPH}.

\begin{algorithm}[h]
\SetAlgoLined
\DontPrintSemicolon
\SetKwInOut{Parameter}{parameter}\SetKwInOut{Input}{input}\SetKwInOut{Output}{output}
\SetKwData{n}{n}\SetKwData{f}{f}\SetKwData{g}{g}
\SetKwData{Low}{l}\SetKwData{x}{x}

\Parameter{a fitting algorithm \texttt{FIT}}
\Input{a density function $\f$, number of phases $\n$}
\Output{a \dCTMC{} $\dctmc = (\states, \events, \sched, \occur, \distribution_0)$}
\BlankLine
\tcp{shift the density function back}
let $\Low > 0$ be the maximal number such that $\f$ is zero on the interval $[0,\Low)$\;
$\g \leftarrow (\x \mapsto \f(\x+\Low))$\;
\BlankLine
$(\states, \events, \sched, \occur, \distribution'_0) \leftarrow \mathtt{FIT}(\n-1,\g)$\tcp*{do the fitting}
\BlankLine
$\states \leftarrow \states \cup \{s\}$\tcp*{add a new initial state}
$\distribution_0(s) \leftarrow 1$
\BlankLine
$\events \leftarrow \events \cup \{d\}$\tcp*{add a discrete-time event scheduled in $s$}
$\sched(s) \leftarrow \{d\}$\;
$\occur(s,d) \leftarrow \distribution'_0$\tcp*{after $d$ occurs, the successor is chosen using $\distribution'_0$}
\caption{\algonep{FIT}}
\label{alg:below}
\end{algorithm}

\begin{algorithm}[h]
\SetAlgoLined
\DontPrintSemicolon
\SetKwInOut{Parameter}{parameter}\SetKwInOut{Input}{input}\SetKwInOut{Output}{output}
\SetKwData{n}{n}\SetKwData{f}{f}\SetKwData{g}{g}
\SetKwData{Up}{u}\SetKwData{x}{x}\SetKwData{s}{s}\SetKwData{Dis}{d}\SetKwData{p}{slices}
\SetKwData{length}{length}\SetKwData{area}{area}\SetKwData{index}{i}\SetKwData{lasts}{previous\_slice}

\Parameter{a fitting algorithm \texttt{FIT}, number of $\p > 1$}
\Input{a density function $\f$, number of phases $\n$ with $\n = k\cdot\p$ for some $k \in \Nset$}
\Output{a \dCTMC{} $\dctmc = (\states, \events, \sched, \occur, \distribution_0)$}
\BlankLine
let $\length < \infty$ be the minimal number such that $f$ is zero on the interval $(\length,\infty)$\;
\BlankLine
\tcp{the first slice}
$(\states, \events, \sched, \occur, \distribution_0) \leftarrow \mathtt{FIT}(\n/\p,\f)$\tcp*{do the fitting}
$\lasts \leftarrow \states$\;
\BlankLine
\tcp{for every other slice}
\For{${\index}\leftarrow 2$ \KwTo $\p$}{
  \tcp{get the conditional density function}
  $\length \leftarrow 1/2 * \length$\tcp*{length of the previous slice}
  $\f \leftarrow (\x \mapsto \f(\x+\length)/(1-F(\length))$ \tcp*{$F$ is the cdf corresponding to $f$}
  \BlankLine
  \tcp{do the fitting}
  $(\states', \events', \sched', \occur', \distribution'_0) \leftarrow \mathtt{FIT}(\n/\p,\f)$\;
\tcp{we assume the sets of states to be disjoint for all slices except for the absorbing state $0$ which we thus add only once
}
  $\states \leftarrow \states \cup \states'$, $\events \leftarrow \events \cup \events'$, $\sched \leftarrow \sched \cup \sched'$, $\occur \leftarrow \occur \cup \occur'$\;
  \BlankLine
  \tcp{connect the previous slice with the current slice}
  $\events \leftarrow \events \cup \{d_{\index-1}\}$\tcp*{by a discrete-time event}
  $delay(d_{\index-1}) \leftarrow \length$\tcp*{of length of the previous slice}
  \For{$s\in\lasts$}{
    $\sched(s) \leftarrow \sched(s) \cup \{d_{\index-1}\}$ \tcp*{scheduled in the whole previous slice}
    $\occur(s,d_{\index-1}) \leftarrow \distribution'_0$\tcp*{leading to the current slice according to $\distribution'_0$}
  }
  $\lasts \leftarrow \states'$\;
}
\caption{\algtwop{FIT,slices}}
\label{alg:above}
\end{algorithm}

%% file: main.bbl
\begin{thebibliography}{10}

\bibitem{AS87Erlang}
D.~Aldous and L.~Shepp.
\newblock The least variable phase type distribution is {E}rlang.
\newblock {\em Communications in Statistics. Stochastic Models}, 3(3):467--473,
  1987.

\bibitem{alur2006bounded}
R.~Alur and M.~Bernadsky.
\newblock Bounded model checking for {GSMP} models of stochastic real-time
  systems.
\newblock In {\em Hybrid Systems: Computation and Control}, pages 19--33.
  Springer, 2006.

\bibitem{asmussen1996fitting}
S.~Asmussen, O.~Nerman, and M.~Olsson.
\newblock Fitting phase-type distributions via the {EM} algorithm.
\newblock {\em Scandinavian Journal of Statistics}, 23(4):419--441, 1996.

\bibitem{baier2003model}
C.~Baier, B.~Haverkort, H.~Hermanns, and J.-P. Katoen.
\newblock Model-checking algorithms for con\-tinuous-time {M}arkov chains.
\newblock {\em Software Engineering, IEEE Trans. on}, 29(6):524--541, 2003.

\bibitem{bernadsky2007symbolic}
M.~Bernadsky and R.~Alur.
\newblock Symbolic analysis for {GSMP} models with one stateful clock.
\newblock In {\em Hybrid Systems: Computation and Control}, pages 90--103.
  Springer, 2007.

\bibitem{bobbio2003acyclic}
A.~Bobbio, A.~Horv{\'a}th, M.~Scarpa, and M.~Telek.
\newblock Acyclic discrete phase type distributions: Properties and a parameter
  estimation algorithm.
\newblock {\em Performance evaluation}, 54(1):1--32, 2003.

\bibitem{bobbio1994benchmark}
A.~Bobbio and M.~Telek.
\newblock {A Benchmark for PH Estimation Algorithms: Results for Acyclic-PH}.
\newblock {\em Communications in statistics. Stochastic models},
  10(3):661--677, 1994.

\bibitem{brazdil2011fixed}
T.~Br{\'a}zdil, J.~Kr{\v{c}}{\'a}l, J.~K{\v{r}}et{\'\i}nsk{\'y}, and
  V.~{\v{R}}eh{\'a}k.
\newblock Fixed-delay events in generalized semi-{M}arkov processes revisited.
\newblock In {\em CONCUR 2011}, pages 140--155. Springer, 2011.

\bibitem{bucci2010oris}
G.~Bucci, L.~Carnevali, L.~Ridi, and E.~Vicario.
\newblock Oris: a tool for modeling, verification and evaluation of real-time
  systems.
\newblock {\em International journal on software tools for technology
  transfer}, 12(5):391--403, 2010.

\bibitem{chen2009quantitative}
T.~Chen, T.~Han, J.-P. Katoen, and A.~Mereacre.
\newblock Quantitative model checking of continuous-time {M}arkov chains
  against timed automata specifications.
\newblock In {\em LICS}, pages 309--318. IEEE, 2009.

\bibitem{cox1955supplementary}
D.~R. Cox.
\newblock The analysis of non-{M}arkovian stochastic processes by the inclusion
  of supplementary variables.
\newblock {\em Math. Proceedings of the Cambridge Phil. Society}, 51:433--441,
  7 1955.

\bibitem{fackrell2005fitting}
M.~Fackrell.
\newblock Fitting with matrix-exponential distributions.
\newblock {\em Stoch. models}, 21(2-3):377--400, 2005.

\bibitem{Faddy98InferringCox}
M.~J. Faddy.
\newblock On inferring the number of phases in a {C}oxian phase-type
  distribution.
\newblock {\em Communications in Statistics. Stochastic Models},
  14(1-2):407--417, 1998.

\bibitem{FW98Heavytail}
A.~Feldmann and W.~Whitt.
\newblock {Fitting Mixtures of Exponentials to Long-Tail Distributions to
  Analyze Network}.
\newblock {\em Perform. Eval.}, 31(3-4):245--279, 1998.

\bibitem{GL:dspn-analysis-supplementary-variables}
R.~German and C.~Lindemann.
\newblock Analysis of stochastic {Petri} nets by the method of supplementary
  variables.
\newblock {\em Performance Evaluation}, 20(1-3):317--335, 1994.

\bibitem{guet2012delayed}
C.~Guet, A.~Gupta, T.~Henzinger, M.~Mateescu, and A.~Sezgin.
\newblock Delayed continuous-time {M}arkov chains for genetic regulatory
  circuits.
\newblock In {\em CAV}, pages 294--309. Springer, 2012.

\bibitem{haas2002stochastic}
P.~J. Haas.
\newblock {\em Stochastic petri nets}.
\newblock Springer, 2002.

\bibitem{haddad2005performance}
S.~Haddad, L.~Mokdad, and P.~Moreaux.
\newblock Performance evaluation of non {M}arkovian stochastic discrete event
  systems -- a new approach.
\newblock In {\em WODES'04}, page 243. Elsevier, 2005.

\bibitem{hatefi2013improving}
H.~Hatefi and H.~Hermanns.
\newblock Improving time bounded reachability computations in interactive
  {M}arkov chains.
\newblock In {\em Fundamentals of Soft. Engineering}, pages 250--266. Springer,
  2013.

\bibitem{horvath2011probabilistic}
A.~Horv{\'a}th, M.~Paolieri, L.~Ridi, and E.~Vicario.
\newblock Probabilistic model checking of non-{M}arkovian models with
  concurrent generally distributed timers.
\newblock In {\em QEST}, pages 131--140. IEEE, 2011.

\bibitem{horvath2012transient}
A.~Horv{\'a}th, M.~Paolieri, L.~Ridi, and E.~Vicario.
\newblock Transient analysis of non-markovian models using stochastic state
  classes.
\newblock {\em Performance Evaluation}, 69(7):315--335, 2012.

\bibitem{horvath2010transient}
A.~Horv{\'a}th, L.~Ridi, and E.~Vicario.
\newblock Transient analysis of generalised semi-{M}arkov processes using
  transient stochastic state classes.
\newblock In {\em QEST}, pages 231--240. IEEE, 2010.

\bibitem{HT-Performance-02}
A.~Horv{\'a}th and M.~Telek.
\newblock {M}arkovian modeling of real data traffic: Heuristic phase type and
  {MAP} fitting of heavy tailed and fractal like samples.
\newblock In {\em Performance}, pages 405--434, 2002.

\bibitem{horvath2002phfit}
A.~Horv{\'a}th and M.~Telek.
\newblock {Phfit: A general phase-type fitting tool}.
\newblock In {\em TOOLS}, pages 1--14. Springer, 2002.

\bibitem{horvath2007momfit}
A.~Horv{\'a}th and M.~Telek.
\newblock Matching more than three moments with acyclic phase type
  distributions.
\newblock {\em Stochastic models}, 23(2):167--194, 2007.

\bibitem{jensen1953markoff}
A.~Jensen.
\newblock Markoff chains as an aid in the study of {M}arkoff processes.
\newblock {\em Scandinavian Actuarial Journal}, 1953(sup1):87--91, 1953.

\bibitem{jones2001phased}
R.~Jones and G.~Ciardo.
\newblock On phased delay stochastic {P}etri nets: Definition and an
  application.
\newblock In {\em Petri Nets and Performance Models}, pages 165--174. IEEE,
  2001.

\bibitem{KNP11}
M.~Kwiatkowska, G.~Norman, and D.~Parker.
\newblock {PRISM} 4.0: Verification of probabilistic real-time systems.
\newblock In {\em CAV}, pages 585--591. Springer, 2011.

\bibitem{LRT:dspnexpress-tool}
C.~Lindemann, A.~Reuys, and A.~Thummler.
\newblock The {DSPNexpress} 2.000 performance and dependability modeling
  environment.
\newblock In {\em Fault-Tolerant Comp.}, pages 228--231. IEEE, 1999.

\bibitem{lindemann1999transient}
C.~Lindemann and A.~Th{\"u}mmler.
\newblock Transient analysis of deterministic and stochastic {P}etri nets with
  concurrent deterministic transitions.
\newblock {\em Performance Evaluation}, 36:35--54, 1999.

\bibitem{marsan1987petri}
M.~A. Marsan and G.~Chiola.
\newblock On {P}etri nets with deterministic and exponentially distributed
  firing times.
\newblock In {\em Advances in Petri Nets}, pages 132--145. Springer, 1987.

\bibitem{molloy1985discrete}
M.~K. Molloy.
\newblock Discrete time stochastic {P}etri nets.
\newblock {\em Software Eng.}, SE-11(4):417--423, 1985.

\bibitem{Neuts81}
M.~F. Neuts.
\newblock {\em Matrix-geometric solutions in stochastic models: an algorithmic
  approach}.
\newblock The Johns Hopkins University Press, Baltimore, 1981.

\bibitem{cinneide1999open_problems}
C.~A. O'Cinneide.
\newblock Phase type distributions: Open problems and a few properties.
\newblock {\em Communications in statistics. Stochastic models},
  15(4):731--757, 1999.

\bibitem{sassoli2007close}
L.~Sassoli and E.~Vicario.
\newblock Close form derivation of state-density functions over {DBM} domains
  in the analysis of non-{M}arkovian models.
\newblock In {\em QEST}, pages 59--68. IEEE, 2007.

\bibitem{silverman1986density}
B.~W. Silverman.
\newblock {\em Density estimation for statistics and data analysis}.
\newblock Monographs on Statistics and Applied Probability. Chapman and Hall,
  1986.

\bibitem{thummler2006gfit}
A.~Thummler, P.~Buchholz, and M.~Telek.
\newblock A novel approach for phase-type fitting with the {EM} algorithm.
\newblock {\em Dependable and Secure Computing}, 3(3):245--258, 2006.

\bibitem{zhang2010model}
L.~Zhang and M.~R. Neuh{\"a}u{\ss}er.
\newblock Model checking interactive {M}arkov chains.
\newblock In {\em TACAS}, pages 53--68. Springer, 2010.

\end{thebibliography}
